\documentclass[twocolumn, trackchanges]{aastex701}

\usepackage{booktabs}
\usepackage{soul}
\usepackage{threeparttable}

\shorttitle{A NICER view of the millisecond pulsar PSR~J2124$-$3358}

\shortauthors{Gonz\'alez-Caniulef~et~al.}

\begin{document}

\title{A NICER view of the millisecond pulsar PSR~J2124$-$3358: evidence for a helium atmosphere}

\author[0000-0001-5848-0180]{Denis Gonz\'alez-Caniulef}
\affiliation{University of Toulouse, CNES, CNRS, IRAP, Toulouse, France}
\email{denis.gonzalez-caniulef@irap.omp.eu}

\author[0000-0002-6449-106X]{Sebastien Guillot}
\affiliation{University of Toulouse, CNES, CNRS, IRAP, Toulouse, France}
\email[show]{sebastien.guillot@utoulouse.fr}

\author[0009-0005-7766-5638]{Pierre Stammler}
\affiliation{University of Toulouse, CNES, CNRS, IRAP, Toulouse, France}
\email{pierre.stammler@irap.omp.eu}

\author[0000-0002-3408-2759]{Lucien Mauviard}
\affiliation{University of Toulouse, CNES, CNRS, IRAP, Toulouse, France}
\email{lucien.mauviard-haag@irap.omp.eu}

\author[0009-0008-3894-4783]{Christine~Kazantsev}
\affiliation{University of Toulouse, CNES, CNRS, IRAP, Toulouse, France}
\email{ckazantsev@irap.omp.eu}

\author[0000-0002-1009-2354]{Anna~L.~Watts}
\affil{Anton Pannekoek Institute for Astronomy, University of Amsterdam, Science Park 904, 1098XH Amsterdam, the Netherlands}
\affiliation{Gravitation and Astroparticle Physics Amsterdam (GRAPPA), University of Amsterdam, 1098XH Amsterdam, The Netherlands}
\email{A.L.Watts@uva.nl}

\author[0000-0002-2651-5286]{Devarshi~Choudhury}
\affil{Anton Pannekoek Institute for Astronomy, University of Amsterdam, Science Park 904, 1098XH Amsterdam, the Netherlands}
\email{d.choudhury@uva.nl}

\author[0000-0002-9407-0733]{Bas~Dorsman}
\affil{Anton Pannekoek Institute for Astronomy, University of Amsterdam, Science Park 904, 1098XH Amsterdam, the Netherlands}
\email{b.dorsman@uva.nl}

\author[0009-0005-8019-0426]{Mariska~Hoogkamer}
\affiliation{Anton Pannekoek Institute for Astronomy, University of Amsterdam, Science Park 904, 1098XH Amsterdam, the Netherlands}
\email{m.m.hoogkamer@uva.nl}

\author[0000-0002-1169-7486]{Daniela~Huppenkothen}
\affiliation{Anton Pannekoek Institute for Astronomy, University of Amsterdam, Science Park 904, 1098XH Amsterdam, the Netherlands}
\email{d.huppenkothen@uva.nl}

\author[0000-0002-0428-8430]{Yves Kini}
\affiliation{Gravitation and Astroparticle Physics Amsterdam (GRAPPA), University of Amsterdam, 1098XH Amsterdam, The Netherlands}
\email{y.kini@uva.nl}

\author[0000-0001-6356-125X ]{Tuomo~Salmi}
\affil{Department of Physics, University of Helsinki, P.O. Box 64, FI-00014 University of Helsinki, Finland}
\email{tuomo.salmi@helsinki.fi}

\author[0000-0002-9049-8716]{Lucas~Guillemot}
\affil{LPC2E, OSUC, Univ Orleans, CNRS, CNES, Observatoire de Paris, F-45071 Orleans, France}
\affil{ORN, Observatoire de Paris, Universit\'e PSL, Univ Orl\'eans, CNRS, 18330 Nan\c{c}ay, France }
\email{lucas.guillemot@cnrs-orleans.fr}

\author[0000-0002-1775-9692]{Isma\"{e}l Cognard}
\affil{LPC2E, OSUC, Univ Orleans, CNRS, CNES, Observatoire de Paris, F-45071 Orleans, France}
\affil{ORN, Observatoire de Paris, Universit\'e PSL, Univ Orl\'eans, CNRS, 18330 Nan\c{c}ay, France }
\email{icognard@cnrs-orleans.fr}

\author[0000-0002-3649-276X]{Gilles~Theureau}
\affil{LPC2E, OSUC, Univ Orleans, CNRS, CNES, Observatoire de Paris, F-45071 Orleans, France}
\affil{ORN, Observatoire de Paris, Universit\'e PSL, Univ Orl\'eans, CNRS, 18330 Nan\c{c}ay, France }
\email{theureau@cnrs-orleans.fr}

\begin{abstract}
Pulse profile modeling has proven to be a powerful technique for determining the mass and radius of neutron stars. To date, this method has been applied to a handful of millisecond pulsars observed by the Neutron Star Interior Composition Explorer (NICER). However, analyses of more millisecond pulsars are necessary to  determine tight constraints on the equation of state of superdense matter. In this study, we present an analysis of the isolated, rotation-powered millisecond pulsar PSR~J2124$-$3358 using the X-ray Pulse Simulation and Inference (X-PSI) package, a publicly available state-of-the-art code for neutron-star relativistic ray tracing and Bayesian parameter inference. 
We use NICER and Chandra observations of this pulsar, exploring different neutron star  
atmospheric compositions and different configurations of the hot polar caps responsible for the pulsed X-ray emission. 
Our analyses favor a helium atmospheric composition, plausibly originating from accretion and subsequent evaporation of a former hydrogen-depleted binary companion. For this composition, and given the faint nature of the source and the low signal-to-noise of the data sets, we obtain broad posterior distributions yielding a  mass $M = 1.8\pm0.5\,M_\odot$ and an equatorial radius $R_{\mathrm{eq}} =  11.7^{+2.6}_{-3.0}$\,km (medians and $68\%$ credible intervals), and infer a configuration consisting of two slightly non-antipodal hot spots. By contrast, when using a hydrogen atmosphere model, the mass and radius decrease by $\sim 0.5\,M_\odot$ and $\sim 1$\,km, respectively. 
Future multiwavelength studies, particularly those incorporating radio and gamma-ray pulse-emission, may provide tighter constraints on the geometry and physical properties of this source.
\end{abstract}



\section{Introduction} 
\label{sec:introduction}

Neutron stars (NSs) are compact astrophysical objects whose core density can exceed several times the nuclear saturation density ($\rho_0=2.8\times10^{14}\mathrm{g\,cm}^{-3}$). 
In particular, the cold, dense matter regime remains largely inaccessible to terrestrial laboratories and presents significant challenges for theoretical modeling from first principles. At such extreme conditions, a variety of compositions have been proposed for the NS interior, ranging from nucleonic matter to more exotic phases, such as hyperons or deconfined quark matter. Accordingly,  one of the main goals in the astrophysics of compact objects is the precise determination of NS properties, such as mass and radius, which can provide information about the interior. Measurements of these quantities are crucial for constraining the equation of state (EOS) of superdense matter, which offers valuable insights into our understanding of fundamental physics \citep[for reviews, see, e.g.,][]{Lattimer2016,Baym2018,Burgio2021,Chatziioannou2025}.

There are several techniques to determine the masses and/or radii of NSs. These include the analysis of radio timing of pulsars in binary systems \citep[e.g.,][]{Arzoumanian2018,Fonseca2021}, pulse profile modeling \citep[PPM, e.g.,][]{Watts2016,Bogdanov2019b}, studies of type-I X-ray bursts on accreting NSs \citep{Steiner2010,Ozel2016,Nattila2017}, and spectral fitting of the cold thermal emission from rotation-powered millisecond pulsars \citep[hereafter MSPs,][]{GonzalezCaniulef2019} or from NSs in quiescent low-mass X-ray binaries \citep[e.g.,][]{Guillot2013,Lattimer2014,Shaw2018,Kazantsev2026}, among others\footnote{For example, the analyses of gravitational waves emitted by double NS mergers constrain the  mass-tidal deformability and have been used to place constraints on the EOS \citep[e.g.,][]{Abbott2017,Abbott2020}}. In particular, PPM has proven especially successful in recent years. This technique has been applied to MSPs using observations from NICER \citep{Gendreau2016}, an X-ray observatory with $\sim 100$~ns timing resolution and spectroscopic capabilities, that has until recently been operational on the International Space Station.

PPM aims to reconstruct the X-ray pulse profile of MSPs, which are produced by X-rays emitted from hot polar caps heated by returning magnetospheric currents \citep[e.g.,][]{Harding2001}. 
A key component of this method is the use of a ray-tracing algorithm, together with atmosphere models,  to describe the paths of photons emerging from the NS surface that travel through the NS space-time and the interstellar medium to finally reach the observer's instrument.
The method incorporates general relativistic effects, such as gravitational light bending and time delay \citep{Pechenick1983}, and special relativistic effects, such as  Doppler shifts and light aberration, including oblateness caused by the rapid rotation of MSPs \citep[e.g.,][]{Miller1998, Morsink2007}. By accounting for these effects, PPM enables  measurements of the mass and radius of NSs, as well as constraints on their hot spot configurations \citep[see e.g.,][and references therein]{Watts2016,Bogdanov2019b}.

To date, PPM has been successfully applied to NICER observations of a handful of MSPs: PSR~J0030$+$0451 \citep{Miller2019,Riley2019,Vinciguerra2024,Kini2026}, PSR~J0740$+$6620 \citep{Miller2021,Riley2021,Salmi2022,Dittmann2024,Salmi2024a}, PSR~J0437$-$4715 \citep{Choudhury2024,Miller2026}, PSR~J1231$-$1411 \citep{Salmi2024b,Qi2025}, and PSR~J0614$-$3329 \citep{Mauviard2025}. The results of these studies have provided typical radii measurements ranging $\sim11-13$~km (some  determined at about $\pm 7\%$ precision at the $68\%$ credible level, e.g., PSR~J0437$-$4715 by \citealt{Choudhury2024}) for masses in the range of $\sim1-2~M_\odot$. 
These measurements are used to statistically infer the EOS 
and any new measurement helps to improve these constraints \citep[see e.g.,][]{Koehn2024,Rutherford2024,Kurkela2024,Golomb2025,Huang2025,Li2025}.  
In addition, the resulting hot spot configurations have been used to investigate  the magnetic field topology of MSPs \citep[see, e.g.,][]{Bilous2019,Chen2020,Kalapotharakos2021,Petri2023,Petri2025,HuangChen2025,Cao2026,Petri2026}. 

An additional candidate for PPM is PSR~J2124$-$3358, an isolated MSP discovered in radio observations with the Parkes telescope \citep{Bailes1997}. Its radio-measured period $P=4.9$\,ms and period derivative $\dot P=2.1\times10^{-20}\,\mathrm{s\,s}^{-1}$ imply a spin-down-derived magnetic field $B=2.3\times10^8$\,G. The source is located at a distance $d = 470\pm20$\,pc, based on the measurement of its timing parallax by the European Pulsar Timing Array \citep[EPTA,][]{EPTA2023}. In X-rays, it was first detected by the ROentgen SATellite (ROSAT), which revealed X-ray emission pulsating at the radio period \citep{Becker1999}. 
Observations with the Chandra X-ray observatory and XMM-Newton have shown that PSR~J2124$-$3358 is i)~embedded in a diffuse pulsar wind nebula (PWN)\footnote{Given the similarity of the spin parameters of PSR~J2124$-$3358 with those of other MSPs (without detected diffuse X-ray emission), \cite{Hui2006} suggested that the presence of a diffuse PWN is likely due to differences in the interaction of the pulsar with the local environment (interstellar medium)  rather than to its total spin-down energy output.} 
and ii)~its spectrum is dominated by thermal emission, with an unabsorbed flux of $1.7\times10^{-13}\,\mathrm{erg\,cm^{-2}\,s^{-1}}$ in the  $0.25-2$~keV range \citep{Hui2006,Zavlin2006,Romani2017}. 
PSR~J2124$-$3358 was the faintest of the four initial MSPs observed by NICER. Preliminary spectral and timing analyses of the source were discussed in \cite{Bogdanov2019a}, based on $\sim 1.6\,\mathrm{Ms}$ NICER observations (performed between 2017 June 26 and 2019 June 30).    

Here, the goal of our work is to perform a full PPM analysis of PSR~J2124$-$3358 using longer NICER observations than those in \cite{Bogdanov2019a}, together with Chandra observations, and to derive mass and radius measurements using the relativistic ray-tracing and Bayesian parameter inference code X-PSI \citep{xpsi2023}. To this end, we investigate various configurations and complexities of the hot polar caps, as well as different atmospheric compositions,  and assess their impact on the resulting mass and radius measurements of the source. 

The paper is organized as follows. In Section~\ref{sec:data}, we present the data reduction of the NICER and Chandra observations of PSR~J2124$-$3358. Section~\ref{sec:PPM} outlines the method used to construct posterior distributions from X-PSI. The results are presented in Section~\ref{sec:results}, followed by the discussion in Section~\ref{sec:discussion} and the conclusions in Section~\ref{sec:conclusions}.

\section{X-ray data} 
\label{sec:data}
\subsection{NICER}
We analyzed the NICER observations of PSR~J2124$-$3358 performed between 2017 June 26 and 2023 May 20. We excluded data affected by the light leak\footnote{\url{https://heasarc.gsfc.nasa.gov/docs/nicer/analysis_threads/light-leak-overview/}} after 2023 May 22, which degraded the performance of the X-ray Timing Instrument (XTI) due to its sensitivity to optical light. Additionally, we omitted observations with ObsIDs 2060040349 and 2060040350, as they were taken during the time stamp anomaly\footnote{\url{https://heasarc.gsfc.nasa.gov/docs/nicer/data_analysis/nicer_analysis_tips.html}} (July 8–22, 2019). 

Data were processed using \textsc{Heasoft} v6.34 and \textsc{NICERsoft}. As a first step, we applied a standard analysis using the \texttt{nicerl2} task (as described in HEASARC webpage\footnote{\url{https://heasarc.gsfc.nasa.gov/docs/nicer/analysis_threads/nicerl2/}}), which takes advantage of the most recent NICER calibration files (CALDB version \texttt{xti20240206}), as well as improved screening and filtering algorithms. We ran this task on all observation segments.

Next, and similar to previous NICER analyses, we used \texttt{psrpipe} to select good time intervals (GTIs) and events according to the following criteria: i)~energy range: 0.22$-$15.1 keV,  ii)~max. overshoot: 1.5 $\mathrm{cts\,s}^{-1}$,  iii)~max. undershoot: 100 $\mathrm{cts\,s}^{-1}$, iv)~min. sun angle: 80$^\circ$, v)~cutoff rigidity  COR\_SAX $> 1.5~\mathrm{GeV\,c}^{-1}$, and vi)~maximum planetary index $K_p$: 5.

We phase-folded each photon using the \texttt{photonphase} task (\texttt{PINT}) based on a timing solution obtained with radio observations from the Nan\c{c}ay Radio Telescope (NRT), using the data and analysis methods presented in \cite{Guillemot2023}.
The radio data cover the full NICER date range. Then, all event files were merged into a single dataset and a light curve was generated in the $2-10$\,keV range, i.e., where no pulsar counts are expected. We explored different count rate cuts for this lightcurve and assessed the resulting pulse profile \citep{Bogdanov2019b,Guillot2019}. We find that 
the largest H-test statistic \citep{deJager1989,deJager2010} for a pulsed signal is obtained for i)~a count rate cut of $1.4~\mathrm{cts\,s}^{-1}$ (in $16$\,s bins), and ii)~ a  selection of events in the
[30,160] PI channel range  (equivalent to $0.3-1.6$\,keV range). After this filtering process, we obtained a final exposure time of 1.72\,Ms (from an initial exposure of 2.28\,Ms).

\begin{figure}[!t] 
\includegraphics[width=\linewidth]{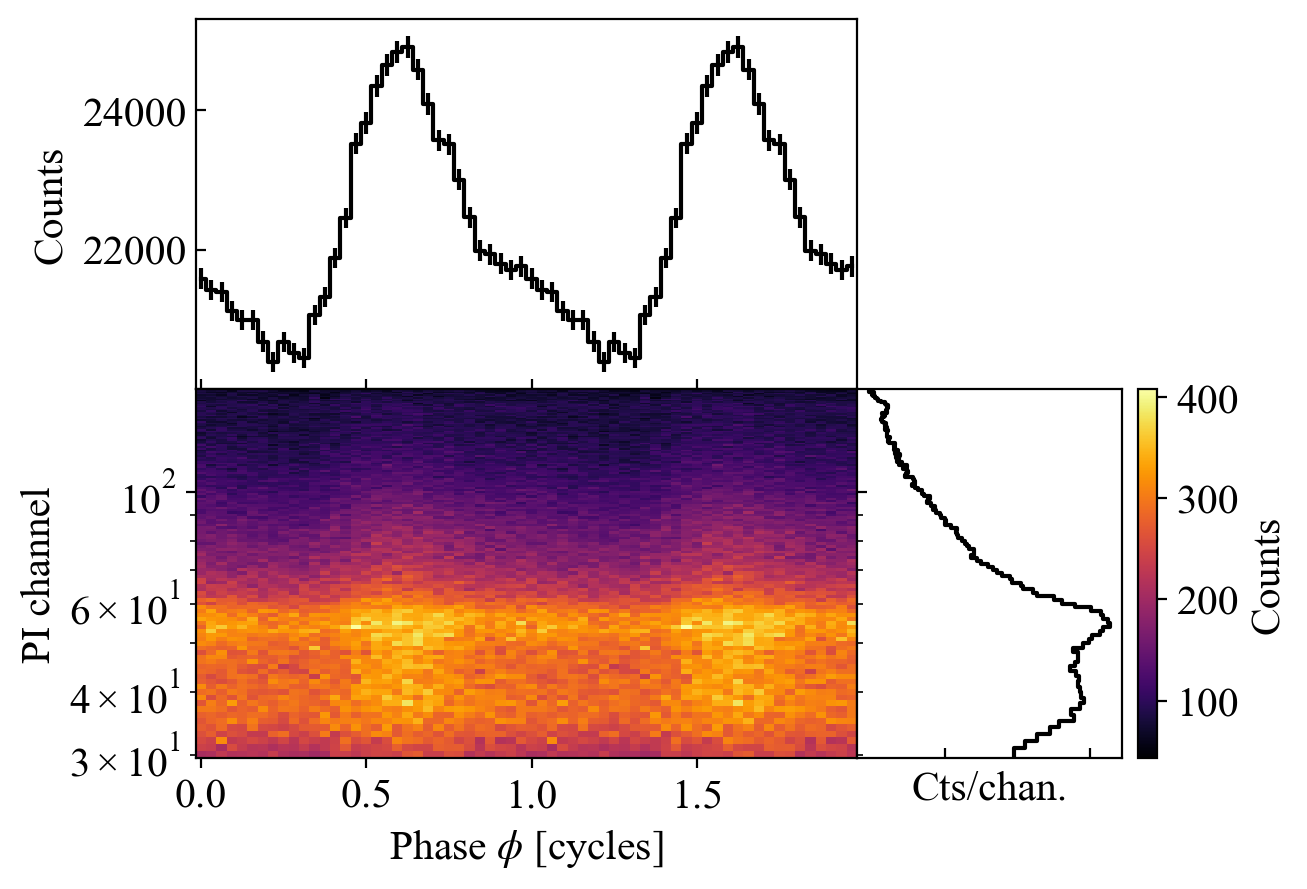}
\caption{Phase-dependent NICER data for PSR~J2124$-$3358 in the $30-160$ PI channel range (equivalent to $0.3-1.6$~keV energy range). The top panel displays the bolometric (energy-integrated) pulse profile with the associated error bars. The bottom panels shows the energy-phase-resolved pulse profile (left) and phase-averaged spectrum in counts per channel (right). The data are divided in 32 equally spaced phase bins. Two rotational cycles are shown for clarity.}
\label{fig:nicer_data}
\end{figure}

As a final step, we produce the ARF and RMF files using the \texttt{nicerl3-spect} task. As discussed in \cite{Bogdanov2019a}, in order to minimize contamination by neighbor X-ray sources, PSR~J2124$-$3358 was observed with an offset pointing of $1'$ \citep[similar to the case of PSR~J0437$-$4715,][]{Choudhury2024}.  We therefore produced the response files considering the source position rather than the XTI pointing. 

Figure \ref{fig:nicer_data} shows the resulting pulse profile obtained with 32 phase bins. It consists of a main peak without a clear secondary peak, but rather a broad bump following the main pulse mostly visible in the bolometric (1D) pulse profile, resembling a mirrored version of PSR~J0437$-$4715, as also previously noted by \cite{Bogdanov2019a}. The phase-averaged spectrum presents a clear phase-independent feature between $50-60$ PI channels, which corresponds to the oxygen VII line (at 574 eV), produced by charge exchange between the solar wind ions and atmospheric atoms within the Earth magnetosphere\footnote{\url{https://heasarc.gsfc.nasa.gov/docs/nicer/analysis_threads/scorpeon-overview/}}. This feature is also present in some recent NICER observations of other MSPs, e.g., PSR~J1231$-$1411 \citep{Salmi2024b},  PSR~J0740$+$6620 \citep{Salmi2024a}, and PSR~J0614$-$3329 \citep{Mauviard2025}.

\subsection{Chandra}

As shown in previous studies \citep[see e.g.,][]{Miller2021,Riley2021,Salmi2024a,Mauviard2025}, PPM of NICER data can greatly benefit from adding information of the source spectrum from other X-ray observations as it helps to indirectly constrain the NICER background (see Section \ref{sec:background}). Following the same approach, we incorporate Chandra observations of PSR~J2124$-$3358 in our analysis, which were previously reported by \citet{Chatterjee2005} and \citet{Romani2017} and retrieved from archival data. These consist of three imaging observations with exposure times of 30, 93 and 85~ks taken with the ACIS-S instrument on 2004 December 19 (ObsID 5585),  2016 July 7 (ObsID 17900) and 2016 September 4 (ObsID 19686), respectively. 
Although XMM–Newton (MOS1/2) observations of PSR~J2124$-$3358 are also available \citep{Hui2006}, the corresponding spectra are significantly contaminated by soft X-ray emission from the surrounding PWN because of the more limited instrumental angular resolution. In contrast, Chandra resolves the extended PWN emission more clearly, leading to a better identification of the source and background spectral extraction regions. Therefore, we restrict our analysis to the Chandra and NICER data.

\begin{figure}[!t] 
\centering
\includegraphics[width=0.9\linewidth]{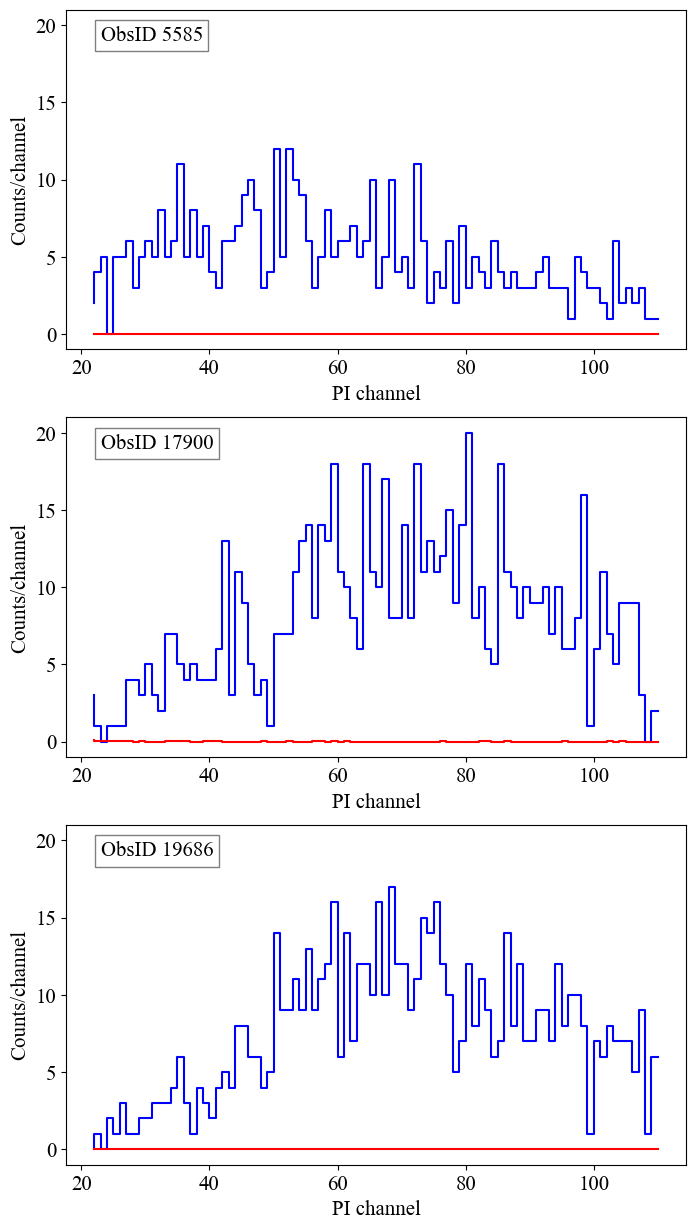}
\caption{Chandra spectral data for PSR~J2124$-$3358 in the $22-110$ PI channel range (equivalent to $0.3-1.6$~keV energy range). The blue and red lines correspond to source and background spectra, respectively. Notice that the background spectra are just slightly above zero.} 
\label{fig:chandra_data}
\end{figure}

We process the data using the task \texttt{chandra\_repro} of the Chandra CIAO package \citep[v. 4.17,][]{Fruscione2006,Fruscione2026}, with the standard options.  We extract the source spectrum considering a circular region of 2" centered on the pulsar and the background from a 25" circular region near the pulsar, devoid of other sources and located away from the PWN. As reported by \cite{Hui2006} and \cite{Romani2017}, pile-up is negligible in these observations ($<0.2\%$). We selected the counts in the $0.3-1.6$~keV range like in the case of NICER data. Figure \ref{fig:chandra_data} shows the resulting source and background spectra for each Chandra ObsID.

\section{Pulse profile modeling}
\label{sec:PPM}
We use X-PSI\footnote{\url{https://xpsi-group.github.io/xpsi/}} \texttt{v3.0.0} and \texttt{v3.1.0} \citep{xpsi2023} to perform PPM of the combined NICER and Chandra observations of PSR~J2124$-$3358. X-PSI is a ray-tracing and  Bayesian inference code designed for modeling the phase- and energy-dependent emission from hot spots, 
as well as from the bulk of the stellar surface
on rotating NSs. It allows for simultaneous modeling of the hot spot geometry and 
atmospheric emission, accounting for special and general relativistic effects in the oblate Schwarzschild approximation \citep[for more details see, e.g.,][]{AlGendy2014,Bogdanov2019b}, while also incorporating  attenuation by the interstellar medium and the effects of the instrumental response of the telescope on the observed radiation.  X-PSI can incorporate data from multiple observatories and employs 
Bayesian sampling algorithms to infer posterior distributions for physical parameters such as the mass, radius, and emission geometry of the NS. In the following, we discuss the X-PSI settings and strategy used to analyze the data of PSR~J2124$-$3358, which closely follows recent X-PSI analyses of other MSPs observed by NICER \citep[e.g.,][]{Salmi2024b,Mauviard2025}.

\subsection{Polar caps}
\label{sec:hotspot}

Motivated by the data structure of the NICER bolometric pulse profile, that is, a main peak followed by a broad bump, we perform X-PSI runs assuming two non-overlapping hot spots. 
As in previous NICER analysis, the modeling of the hot spots is  motivated by
theoretical  works on NS magnetospheres, which suggest the formation of complex returning currents, which in turn may lead to the formation of more complex hot regions than circular spots with uniform temperature \citep[e.g.,][]{Harding2001,Harding2011,Timokhin2013,Kalapotharakos2014,Gralla2017,Lockhart2019,Kalapotharakos2021}. We explore different levels of surface-pattern complexity and  use the following definitions to describe them (as in \citealt{Riley2019}; see also Figure 1 in \citealt{Vinciguerra2023} for a schematic representation), 

\begin{itemize}
\item[1.] \texttt{ST-U}: Single-temperature (\texttt{ST})  regions with unshared (\texttt{U}) parameters. These correspond to two  non-overlapping, spherical regions, each of them with different sizes, locations, and (uniform) temperatures. The primary spot is identified with the lowest colatitude with respect to the spin axis. 
\item[2.] \texttt{ST+CDT}: A \texttt{ST} region plus a Concentric Double Temperature (\texttt{CDT}) region. In this configuration, the secondary spot has extra complexity by adding a centered annulus emitting at a different temperature. The central region and annulus are identified in X-PSI as superceding and ceding regions, respectively.    
\item[3.] \texttt{ST-U+EL}: A \texttt{ST-U} configuration plus an \texttt{elsewhere} component. In this case, \texttt{elsewhere} corresponds to the rest of the NS surface, emitting with uniform temperature. 
\end{itemize}

Table \ref{tab:params} summarizes the parameters that determine the geometry and complexity of a single hot spot, as well as other parameters discussed in the following subsections. 

While other hot spot complexities are also available in X-PSI, such as the protruding double temperature (\texttt{PDT}) region, essentially corresponding to a hot spot surrounded by a possible protruding off-centered annulus, the number of free parameters also increases. 
As discussed in Section~\ref{sec:results}, model performance of \texttt{ST+CDT} seemed satisfactory, however, the constraints are weak, and the credible intervals are wide.
We therefore did not expect any benefit from running a more computationally costly model for this data set.

The addition of the \texttt{elsewhere} component is motivated by a potential colder emission from the bulk of the NS surface, which may contribute in the NICER band. All these considerations are discussed in more detail in Section \ref{sec:discussion}. 

\begin{deluxetable}{ll}[t]
\label{tab:params}
\tablecaption{Model Parameters}
\tablehead{\colhead{\bf{Symbol}} & \colhead{\bf{Meaning}}} 
\startdata
$M\,[\mathrm{M_\odot}]$ & Gravitational mass of the MSP\\
$R_{\mathrm{eq}}\,[\mathrm{km}]$ & Coordinate equatorial radius of the MSP \\
$D\,[\mathrm{kpc}]$ & Distance to the MSP\\
$i\,[\mathrm{rad}]$ & Angle between the spin axis and line-of-sight\\
$N_{\mathrm{H}}\,[\mathrm{cm^{-2}}]$ & Neutral hydrogen column density\\
\hline
\multicolumn{2}{c}{Energy-independent effective area scaling factors}\\
\hline
$\alpha_{\mathrm{NICER}}$ & NICER XTI effective area scaling\\
$\alpha_{\mathrm{ACIS1}}$ & Chandra ACIS ObsID 5585 scaling \\
$\alpha_{\mathrm{ACIS2}}$ & Chandra ACIS ObsID 17900 scaling\\
$\alpha_{\mathrm{ACIS3}}$ & Chandra ACIS ObsID 19686 scaling\\
\hline
\multicolumn{2}{c}{Primary or secondary hot spot$^\dag$}\\
\hline
$T_p[\mathrm{K}]$ &  Hot spot temperature\\
$\zeta_p \,[\mathrm{rad}]$ & Angular radius of the hot spot\\
$\theta_p \,[\mathrm{rad}]$ & Colatitude of the hot spot\\
$\phi_p \,[\mathrm{cycles}]$ & Phase offset of the hot spot\\
\hline
\multicolumn{2}{c}{Ceding region of the  secondary spot (if present$^\dag$$^\dag$)}\\
\hline
$T_{c,s}[\mathrm{K}]$ &  Ceding-region temperature\\
$\zeta_{c,s} \,[\mathrm{rad}]$ & Angular radius of the ceding region\\
\hline
\multicolumn{2}{c}{\texttt{Elsewhere} component (if present)}\\
\hline
$T_{\mathrm{else}}[\mathrm{K}]$ &  Temperature of the rest of the surface
\enddata
\tablecomments{$^\dag$ For the secondary hot spot, the $p$ subscript is replaced with a $s$ subscript.\\$^\dag$$^\dag$ Applicable in the case of a \texttt{CDT} hot spot instead of \texttt{ST} (see Section \ref{sec:hotspot}).}
\end{deluxetable}

\subsection{Atmosphere model}

The thermal emission from the hot spots of MSPs is expected to be reprocessed by an atmosphere, whose modeling depends on several assumptions. Due to strong gravitational fields, NS atmospheres are thought to be stratified, with light elements on top of the atmosphere and heavy elements sinking to the deepest layers \citep{Romani1987}. A light element composition may result from past accretion onto the MSP from its binary companion during the recycling phase or accretion of material from the interstellar medium. Furthermore, while NSs can harbor strong magnetic fields, in the case of MSPs these fields ($\sim10^8-10^9$~G) are weak enough so that they do not affect the radiative transfer calculations in the X-ray band \citep[e.g.,][and references therein]{Potekhin2014}. Although another effect, namely particle bombardment from the returning magnetic currents, is important in generating the hot spots, energy deposition occurs in the deepest layers of the atmosphere, making the expected thermal emission similar to that of a passively cooling NS \citep{Bogdanov2007,Baubock2019,Salmi2020}. For the typical temperature of MSP hot spots, $\sim10^6$~K, light-element atmospheres are also expected to be fully ionized. Therefore, and as in previous X-PSI analyses, we use non-magnetic, fully-ionized, hydrogen atmosphere models precomputed with the NSX code \citep{Ho2001,Ho2009}.  We also consider helium atmosphere models, as PPM has been shown to be sensitive to the chemical composition \citep[e.g.,][and references therein]{Salmi2023}.

\subsection{Prior distributions}

PSR~J2124$-$3358 is an isolated MSP, which means that its mass cannot be derived from radio observations contrary to other MSPs belonging to binary systems, e.g., PSR~J0437$-$4715 \citep{Choudhury2024} or PSR~J0614$-$3329 \citep{Mauviard2025}. Therefore, we consider a broad uniform prior for the mass with boundaries $[1.0,3.0]\,M_\odot$, such as for previous analyses of PSR~J0030$+$0451.

As in previous analyses, we apply a uniform prior distribution for the radius within the boundaries $[3 r_\mathrm{g}(1.0\,M_\odot),16.0]$\,km, where $r_\mathrm{g}(M)=GM/c^2$ is the gravitational radius. The radius and mass priors are further modified by  the compactness condition $R_{\mathrm{polar}}/r_\mathrm{g}(M)>3$ ($R_{\mathrm{polar}}$ is the polar radius), and the precomputed surface gravity grid for the atmosphere models discussed below. 

The available NSX atmosphere tables for hydrogen and helium are precomputed over different ranges of effective temperature and surface gravity. In order to facilitate the  comparison of the Bayesian evidence and ranking of the X-PSI models with different atmospheric compositions, we adopt a uniform prior in temperature with $\log_{10} T_\mathrm{eff}\in [5.1,6.5]$ and in surface gravity with $\log_{10} g_\mathrm{eff} \in [13.7,14.7]$, which are the intervals common to both the hydrogen and helium NSX tables.

For the distance prior we adopt a normal distribution centered at $\mu_d=470$~pc with a standard deviation $\sigma_d=20$~pc, and truncated at $\pm 3\sigma_d$, consistent with the latest measurement for PSR~J2124$-$3358 reported by the EPTA collaboration \citep{EPTA2023}. This distance is derived from  parallax measurement obtained with $\sim 16$~years monitoring of the source with radio telescopes such as the NRT and Lovell Telescope (Jodrell Bank Observatory).  

In addition, for the neutral hydrogen column density,  a preliminary analysis of the NICER data by \cite{Bogdanov2019a} derived $N_\mathrm{H}=(3.1\pm0.8)\times10^{20}~\mathrm{cm}^{-2}$ in the direction to the source. However, this estimate was obtained assuming fixed values for the mass and radius, adopting a lower distance for PSR~J2124$-$3358, and with a poor knowledge of the NICER background modeling at the time. Given the uncertainties,  in the following, we just consider a uniform prior for $N_\mathrm{H}$ within the boundaries $[0.001, 20.0]\times10^{20}~\mathrm{cm}^{-2}$. The total interstellar absorption is computed using the \texttt{tbnew}\footnote{\url{https://pulsar.sternwarte.uni-erlangen.de/wilms/research/tbabs/}} model, an improved version of the X-ray absorption model \texttt{tbabs} \citep{Wilms2000}.

The calibration uncertainties in the effective area of the NICER and Chandra instruments are  accounted for in X-PSI by energy-independent effective-area scaling-factors. For our analysis, the NICER XTI instrument is identified by $\alpha_\mathrm{NICER}$, and for Chandra ACIS-S ObsIDs 5585, 17900, and 19686 they are  $\alpha_\mathrm{ACIS1}$, $\alpha_\mathrm{ACIS2}$, and $\alpha_\mathrm{ACIS3}$, respectively. For the XTI scaling factor, and similar to previous NICER analyzes, we consider a prior modeled as a normal distribution centered at $\mu_\alpha=1.0$ with a standard deviation $\sigma_\alpha=10\%$, truncated  at $\pm3\sigma_\alpha$. Similarly, for the ACIS scaling factors, we implement priors as normal distributions, but with a tighter standard deviation of $3\%$, reflecting the better-characterized calibration uncertainties of the Chandra instrument.

\subsection{Background}
\label{sec:background}
As discussed in the preliminary analysis by \cite{Bogdanov2019a},  PSR~J2124$-$3358 is surrounded by an X-ray emitting diffuse PWN, and several X-ray sources are also in the NICER field of view ($\sim 30~\rm{arcmin}^2$), all of which contribute to the background. In addition, NICER observations are well known to be significantly affected by optical loading (optical photons incident on the detectors originating mainly from the Sun), energetic particles, and solar wind charge exchange (X-ray radiation due to the interaction of the solar wind and atmospheric atoms within Earth's magnetosphere). In fact, \cite{Bogdanov2019a} estimated that these combined components can contribute up to $\sim80\%$ of the total NICER counts of PSR~J2124$-$3358 in the relevant energy range. Therefore, a proper characterization of the background is crucial to avoid biasing the PPM results. 

In the following, we configure X-PSI to constrain the NICER background via a joint fitting that includes phase-averaged data taken from Chandra observations. The high-resolution imaging capabilities of Chandra provide a clean estimate of the source spectrum, which is used to constrain the NICER spectrum, and thus indirectly the NICER background. In particular, X-PSI performs a background marginalization procedure (for details, see Appendix B.2 in \citealt{Riley2019PhDT}; \citealt{Riley2021}; \citealt{Salmi2022}). This is applied to each Chandra ObsID, allowing their background to vary uniformly between $\texttt{max}(0, c_b - n \sqrt{c_b})$ and $c_b + n \sqrt{c_b}$, where $c_b$ corresponds to the Chandra background counts per channel and  where we set a tolerance $n=3$ \citep[like in the recent analysis by][]{Mauviard2025}. A discussion on alternative ways to constrain the NICER background using background models, such as 3C50 \citep{Remillard2022}, can be found in \cite{Salmi2022} and \cite{Choudhury2024}.

\subsection{Model resolution}

For the parameters that characterize the surface patterns, the strategy of recent NICER analyses involves the use of low-resolution (LR) settings for exploratory runs and high-resolution (HR) settings for the headline result.
Noting that previous X-PSI analyses may have used slightly different settings, we use the following LR settings for the hot spots \citep[as in][]{Mauviard2025}: 
\texttt{sqrt-num-cells}=16,  
\texttt{min-sqrt-num-cells}=16,  
\texttt{max-sqrt-num-cells}=16,  
\texttt{num-leaves}=32,  
\texttt{num-rays}=100,   
\texttt{num-energies}=64,
and \texttt{image-order-limit}=3.
If an \texttt{elsewhere} component is present, then the LR settings for this component are
\texttt{elsewhere-sqrt-num-cells}=32 and
\texttt{elsewhere-num-rays}=200.
The definitions of these parameters can be found in the X-PSI documentation\footnote{\url{https://xpsi-group.github.io/xpsi/hotregion.html}} \citep[for illustrative HR settings see, e.g.,][]{Vinciguerra2023}.

As discussed in the next sections, our preferred models using LR settings turn out to be computationally expensive, and  HR runs are thus prohibitive given the computational resources available to us. However, \cite{Vinciguerra2023,Vinciguerra2024} have shown that the adoption of LR or HR settings does not have a significant impact on the posterior distributions obtained from synthetic or real data (at least for the case of PSR~J0030$+$0451). Therefore, we limited our analysis only to LR runs, leaving the more computationally expensive HR runs for future investigations \citep[for detailed studies of the effects of resolution on the modeled waveform, see][]{Choudhury2024b}. For other sources, a recent discussion about the computational cost and  improvement in likelihood evaluation when comparing LR and HR models to the data can be found in \cite{Choudhury2024} and \cite{Mauviard2025}.

\subsection{Sampling}
To finalize with the  X-PSI settings, the Bayesian analysis and computation of the posterior distributions are performed with \textsc{PYMULTINEST} \citep{Buchner2014}. This is a \texttt{PYTHON} wrapper of \textsc{MULTINEST}
\citep{Feroz2009}, a nested sampling algorithm \citep{Skilling2004} that is capable of efficiently exploring parameter space, dealing with complex likelihood shapes, and computing model evidence. For our 
runs, we used multimode  exploration turned-off, with sampling efficiency fixed to 0.1, and live points set to  $5\times10^3$ for the simplest model (\texttt{ST-U})  and $10^4$ for the most complex models (\texttt{ST+CDT} and \texttt{ST-U+EL}). 

As discussed, for example, in \cite{Vinciguerra2023}, a low sampling efficiency combined with a high number of live points leads to improved accuracy and precision in both the Bayesian evidence estimates and posterior distributions. Consistently, our  settings are on the conservative and computationally expensive side in order to mitigate the impact of the randomness of the sampling process inherent to \textsc{MULTINEST} and to guarantee an adequate exploration of the model parameter space (see, e.g., \citealt{Vinciguerra2024,Kini2026} for thorough investigations of the effects of \textsc{MULTINEST} settings in the context of X-PSI analyses of PSR~J0030$+$0451, \citealt{Salmi2024a}  for the case of PSR J0740$+$6620, and \citealt{Hoogkamer2025} for a comparison with \textsc{UltraNest} \citep{Buchner2021}).

\begin{figure*}[!t]
    \centering
    \includegraphics[width=0.49\linewidth]{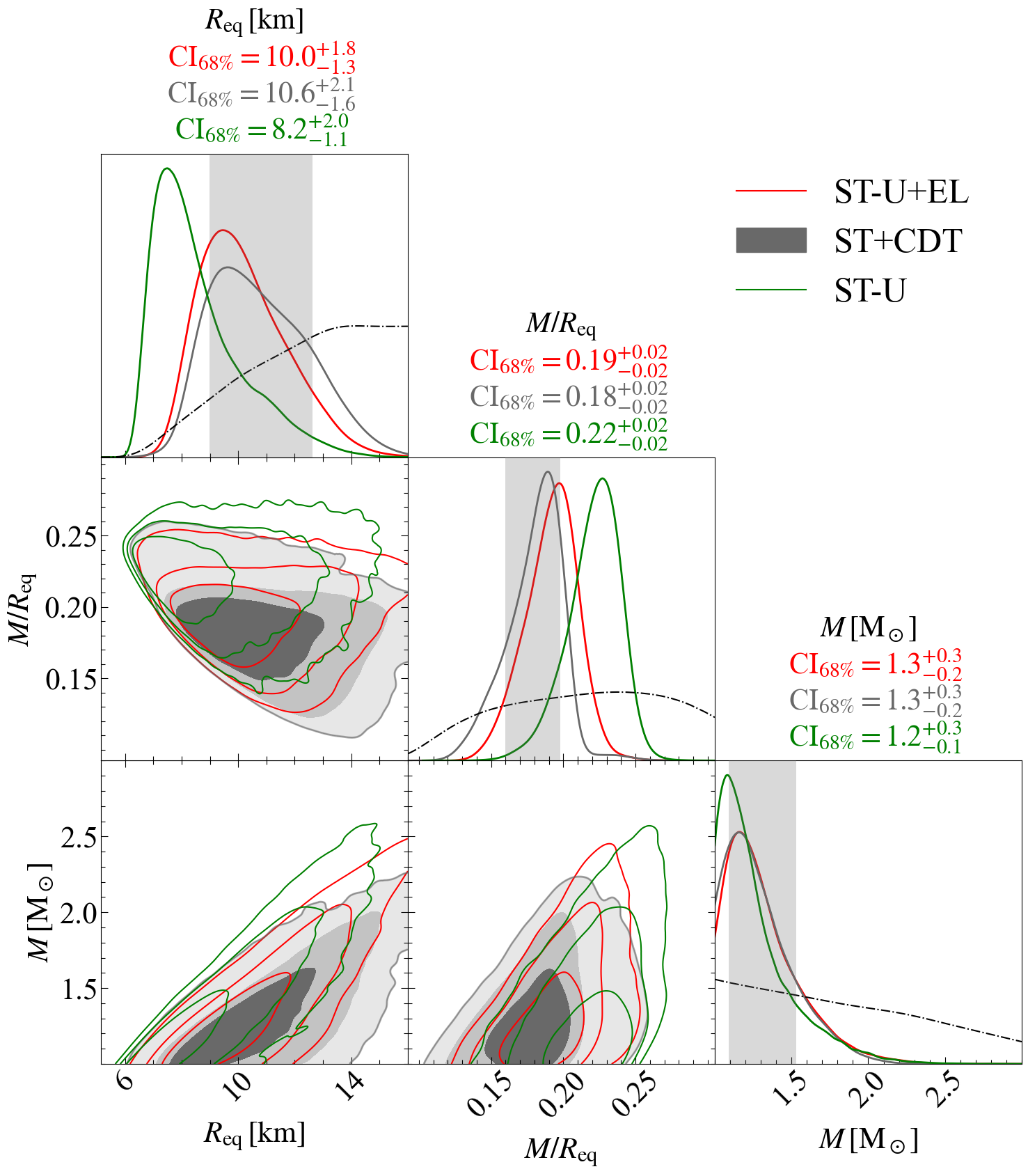}
    \includegraphics[width=0.49\linewidth]{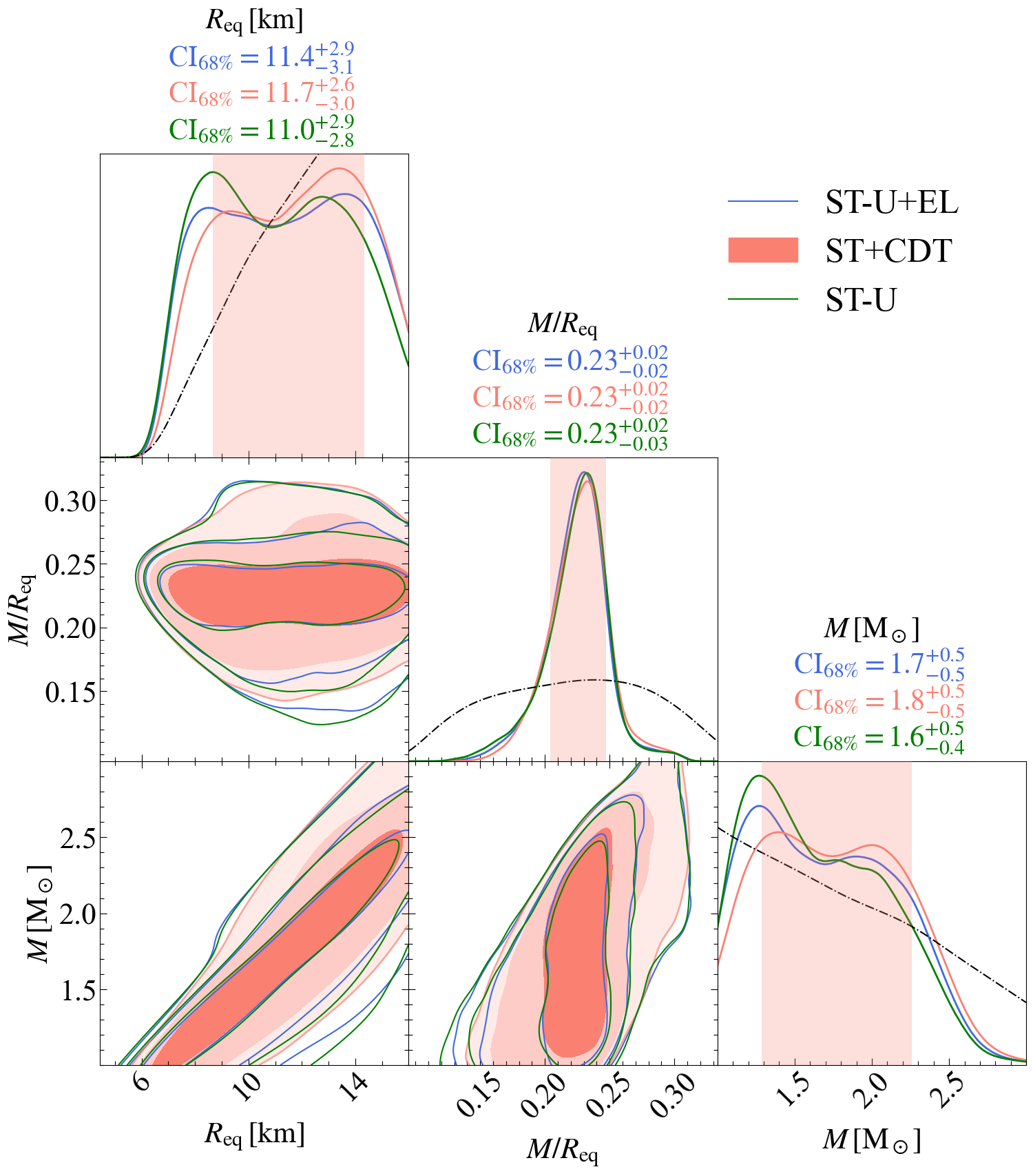}
    \caption{Posterior distributions for the radius, compactness, and mass considering hot spots with hydrogen (left panel) and helium (right panel) atmospheric compositions. For a given atmospheric composition, different colors correspond to different hot spot complexities (\texttt{ST-U}, \texttt{ST+CDT}, or \texttt{ST-U+EL} model). In both panels, the dash-dotted black lines show the 1D  prior distributions, which are common for all models. The shaded, vertical bands (gray in the left panel and salmon in the right panel) show the $68.3\%$ credible intervals, corresponding only to the \texttt{ST+CDT} model. The 2D posterior distributions show the $68.3\%$, $95.4\%$, and $99.7\%$ credible regions. }
    \label{fig:combined_posteriors}
\end{figure*}

\section{Results}
\label{sec:results}

Since the main science goal of NICER is the estimation of the masses and radii of NSs, we first present the results for these parameters, including the compactness\footnote{In the following, the compactness is calculated as $GM/R_{\mathrm{eq}}c^2$, assuming $G=c=1$.}, to which the analysis is expected to be most sensitive. We analyze separately the cases of hydrogen and helium atmospheric compositions, leaving the details of the hot-spot geometry to the subsequent section.
All the results of our PPM (samples, reproduction and post processing scripts) can be found in a Zenodo repository\footnote{\url{https://doi.org/10.5281/zenodo.20640366}}.

\subsection{Hydrogen atmosphere}

We performed X-PSI runs, initially assuming that the NS surface has a hydrogen atmospheric composition. 
Figure \ref{fig:combined_posteriors}, left panel, shows the posterior distributions for the mass, radius, and compactness obtained for the \texttt{ST-U}, \texttt{ST+CDT}, and \texttt{ST-U+EL} models.  In particular, the inferred radii show a substantial dependence on the complexity of the surface pattern. 
The median values are relatively small, $\lesssim 10.6$\,km, particularly when compared to those derived from other NS observations and EOS constraints (based on dense-matter models), which typically suggest $R\approx12$~km \citep[see e.g., for a single EOS model family,][]{Raaijmakers2021,Rutherford2024}.
Nevertheless, they are still consistent with previous radius constraints for this source $>7.8$\,km ($68\%$ confidence) derived using XMM-Newton data and hydrogen atmosphere models \citep{Bogdanov2007}, with a substantial fraction of the posteriors lying above this limit.
Furthermore, we obtain broad radius posteriors for all models, yielding $68\%$ credible intervals as large as $\sim2$\,km in the case of the \texttt{ST+CDT} model. 
This is not surprising, as PSR~J2124$-$3358 is a faint  X-ray source and there is no informative mass or inclination prior.
The posteriors do however show a noticeable shift from the priors, indicating that the data still provide interesting constraints.

\begin{table*}[!ht]
    \centering
    \begin{tabular}{lccccc}
        \hline
        \hline
         \multicolumn{1}{c}{} & \multicolumn{2}{c}{Hydrogen}  & \multicolumn{2}{c}{Helium} & \\
         \cmidrule(lr){2-3} \cmidrule(lr){4-5}
        Model & $\ln \mathcal{Z}$ & $\ln p( \text{d} | \theta_\mathrm{ML})$  & $\ln \mathcal{Z}$ & $\ln p( \text{d} | \theta_\mathrm{ML})$ & $\Delta \log_{10} \mathcal{Z}$\\
         \hline
         \texttt{ST-U} & $-22116.023\pm0.084$ & $-22075.339$ & $-22110.648\pm0.081$ & $-22072.029$ & 2.334\\ 
        \texttt{ST+CDT} & $-22110.530\pm0.061$ & $-22065.945$ &  $-22108.579\pm0.056$ & $-22070.961$ & 0.847\\ 
         \texttt{ST-U+EL} & $-22113.467\pm0.063$ & $-22067.668$ & $-22110.855\pm0.058$ & $-22069.632$ & 1.134\\
         \hline
    \end{tabular}
    \caption{Bayesian evidence and maximum likelihood values for the different hot spot models and atmospheric composition obtained with \textsc{PYMULTINEST} and X-PSI. The $\Delta \log_{10} \mathcal{Z}$ values correspond to the logarithmic difference in evidence between helium and hydrogen composition for a given hot spot model, which allow direct interpretation according to the criteria of \cite{Kass1995}.}
    \label{tab:evidence}
\end{table*}

On the other hand, the inferred masses are fairly similar between models, yielding
median values of $\sim 1.3\,M_\odot$, which is fully consistent with measured masses of NSs in binary systems\footnote{\url{https://www3.mpifr-bonn.mpg.de/staff/pfreire/NS_masses.html}} (noting that, although isolated, PSR J1231$-$1411 must have been part of a binary system to become an MSP, see also Section \ref{sec:discussion}). 
However, we note a caveat that the peaks of the mass posteriors for all models lie close to the lower boundary of the prior \citep[$1\,M_\odot$, approximately consistent
with the validity\footnote{The validity actually depends on the compactness, which for values $\lesssim 0.15$ begins to deviate from the oblateness derived from realistic EOSs. Note that there is also a dependency on the spin frequency of the NS. Initial tests were performed for a different source at 271.45\,Hz (PSR J1231$-$1411), while the frequency of PSR J2124$-$3358 is 202.79\,Hz.} of the oblateness approximation used in X-PSI,][]{AlGendy2014}. 
As a result, the posterior distributions are truncated, and the reported mass estimates may be influenced by the prior lower-mass bound.
Since the radius is correlated with the mass, the inferred radius distributions may also be affected. 
In addition, there is some scatter in the compactness, which is in any case small and for all models consistent within the $68\%$ credible intervals.
The 2D residuals for all surface patterns (available in Zenodo; see also Figure~\ref{fig:residuals} in Appendix A for the case of \texttt{ST+CDT} model)
are similar and do not show significant structures, indicating that all models can reproduce the data relatively well. 

Table~\ref{tab:evidence} shows a summary of the Bayesian evidence obtained for the different model complexities. By computing the base-10 logarithmic differences in evidence and using the \texttt{ST-U} model as the reference, which allows direct comparison according to the Bayesian criteria of \cite{Kass1995},  we find that the \texttt{ST+CDT} model is decisively favored ($\Delta\log_{10}\mathcal{Z}$=2.386), and the \texttt{ST-U+EL} model is strongly favored ($\Delta\log_{10}\mathcal{Z}$=1.110). This implies that the \texttt{ST+CDT} is the preferred model for hydrogen atmospheric composition, which yields $R_{\mathrm{eq}}=10.6^{+2.1}_{-1.6}$\,km and $M=1.3^{+0.3}_{-0.2}\,M_\odot$.

\subsection{Helium atmosphere}

We now present the results of our PPM assuming a NS surface with helium atmospheric composition. Figure \ref{fig:combined_posteriors}, right panel, shows the posterior distributions for the radius, mass, and compactness for the \texttt{ST-U}, \texttt{ST+CDT}, and \texttt{ST-U+EL} models. Compared with the hydrogen runs, some significant differences emerge. For example, by varying the complexity of the model, the measured radius shows relatively small variations (as opposed to the results for hydrogen composition).  The median values of the measured radii are now substantially larger, with all models yielding radii above $R_\mathrm{eq}\sim11$\,km. However, the posterior also presents a mild bimodality, which also makes the $68\%$ credible intervals larger ($\approx \pm 3$~km). 

Similarly, the masses are also higher, with typical values of $\sim1.7~M_\odot$. A mild bimodality is also present in the posteriors, which makes the credible intervals also increase compared to mass measurements of the hydrogen case. The mass and radius are correlated, indicating that an independent measurement of the mass would help  constraining the radius and vice-versa.  
For all models, we obtain remarkably identical median values for the compactness ($M/R_{\mathrm{eq}}=0.23$) with just small differences in the associated $68\%$ credible intervals. 
Again, as in the case of hydrogen, all models can reproduce the data similarly well, and the 2D residuals (available in Zenodo; see also Figure~\ref{fig:residuals} in Appendix A for the case of \texttt{ST+CDT} model) show no significant structures. 

By comparing the Bayesian evidence shown in Table \ref{tab:evidence} and considering the \texttt{ST-U} model as reference, we find that the \texttt{ST+CDT} model is substantially favored according to the criteria of \citet{Kass1995}, with $\Delta\log_{10}\mathcal{Z}=0.899$. Instead, the \texttt{ST-U+EL} model performs quite similar to  the \texttt{ST-U} model with $\Delta\log_{10}\mathcal{Z}=-0.090$, which is a  difference in evidence that is not worth more than a bare mention. This means that the \texttt{ST+CDT} is the preferred model for helium atmospheric composition, which yields $R_{\mathrm{eq}}=11.7^{+2.6}_{-3.0}$\,km and $M=1.8\pm0.5\,M_\odot$.

\begin{figure*}[!ht]
    \centering
    \includegraphics[width=0.8\linewidth]{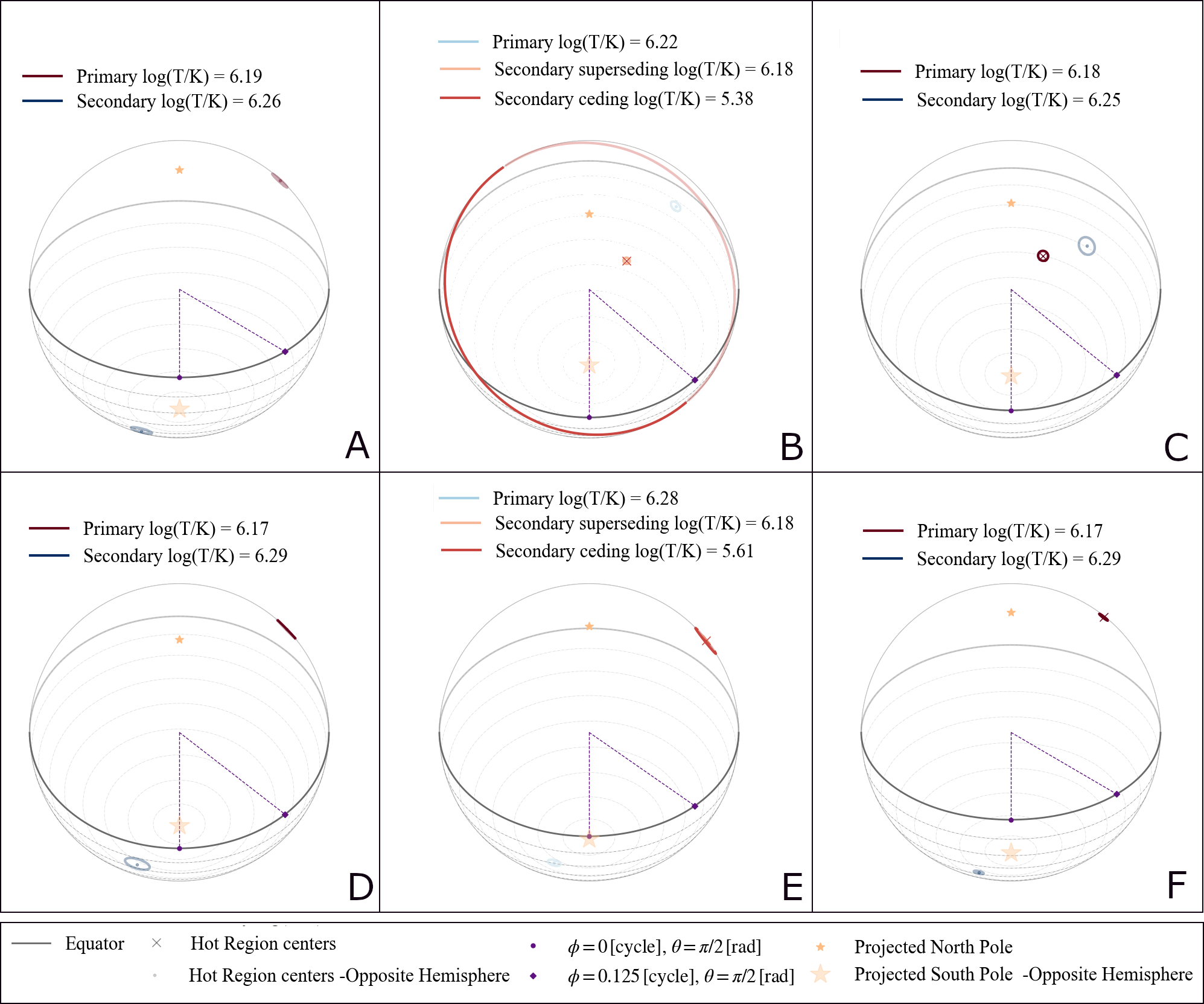}
    \caption{Representation of the geometries associated with the maximum likelihood parameters for different atmosphere compositions and surface pattern complexities. The upper panels (A), (B), and (C) correspond to hydrogen atmospheric composition for the \texttt{ST-U}, \texttt{ST+CDT}, and \texttt{ST-U+EL} models, respectively. The lower panels (D), (E), and (F) show again the \texttt{ST-U}, \texttt{ST+CDT}, and \texttt{ST-U+EL} models, respectively, but assuming helium atmospheric composition. The viewing geometries correspond to phase zero of the NS rotation as observed from Earth. The NS is represented without including general or special relativistic effects. In the panels (B) and (E),  blue, orange, and red colors represent relatively hot, warm, and cold temperatures, respectively, while dimmer colors indicate location of the spot (or part of it) behind the sphere. A similar color scheme is adopted for panels (A), (C), (D), and (F), but using slightly darker tones. In panels (C) and (F), the temperature of the rest of the surface is $\log_{10}(T_{\rm{else}}/\rm{K})=5.46$ and  $5.37$ for the case of hydrogen and helium, respectively.  }
    \label{fig:geometries}
\end{figure*}

\subsection{Hydrogen or helium?}

By comparing the Bayesian evidence for the same hot spot model between hydrogen and helium atmosphere models (see Table \ref{tab:evidence}), we find that  the helium composition is preferred in all cases. 
Furthermore, the median values of the radii obtained with a helium composition are in better agreement  with  both the radii inferred from previous NICER analyses with X-PSI and radii inferred for particular EOS model (incorporating astrophysical constraints and nuclear physics inputs). 
These results support adopting the helium atmospheric composition for the headline result. However, for the preferred \texttt{ST+CDT} model, the difference in Bayesian evidence $\Delta\log_{10}\mathcal{Z}=0.847$ implies that helium composition is substantially, but not decisively, favored according to the criteria of \citet{Kass1995}. Therefore, in the following, the \texttt{ST+CDT} configuration for hydrogen is also discussed, as the data do not rule out this model and the residuals are similar to the case of helium composition
(see Figure~\ref{fig:residuals} in Appendix A). 

It is interesting to note that the \texttt{ST+CDT} model with LR settings required $\sim 31$ and $\sim221$~khrs of CPU time, in the case of hydrogen and helium atmospheric composition, respectively. 
Given our computational constraints and the fact that the credible intervals are wide, we therefore do not consider more expensive HR runs for this work. 

\begin{figure*}[!ht]
    \centering
    \includegraphics[width=0.49\linewidth]{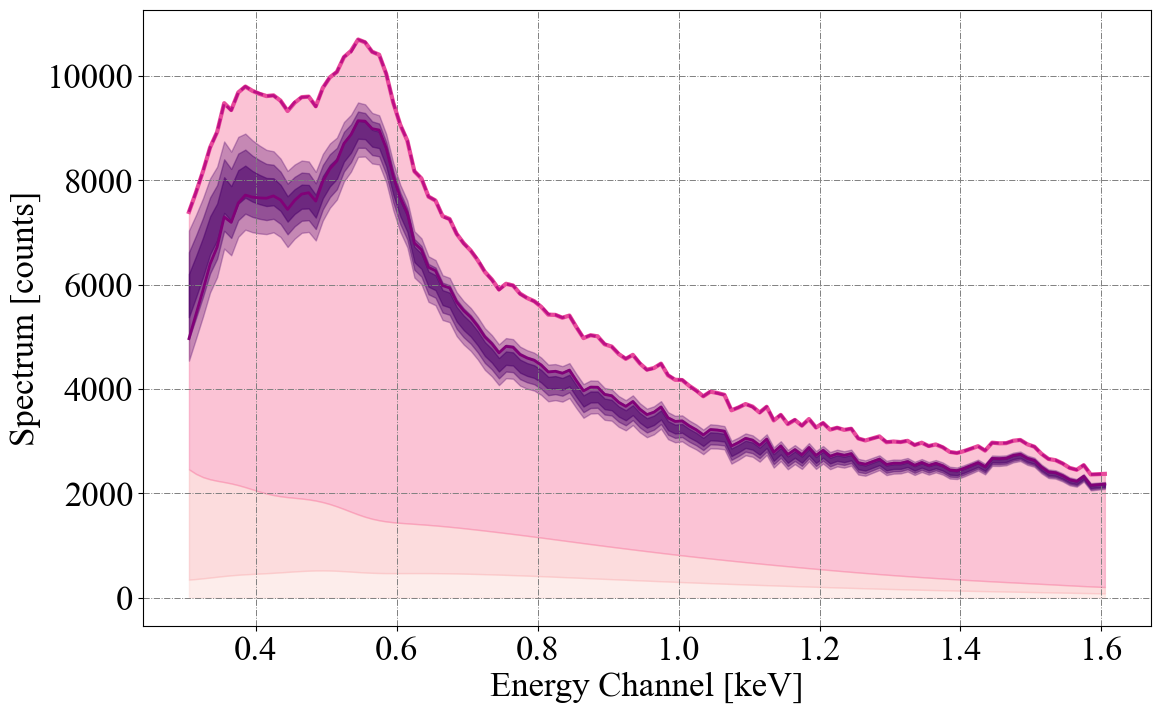}
    \includegraphics[width=0.49\linewidth]{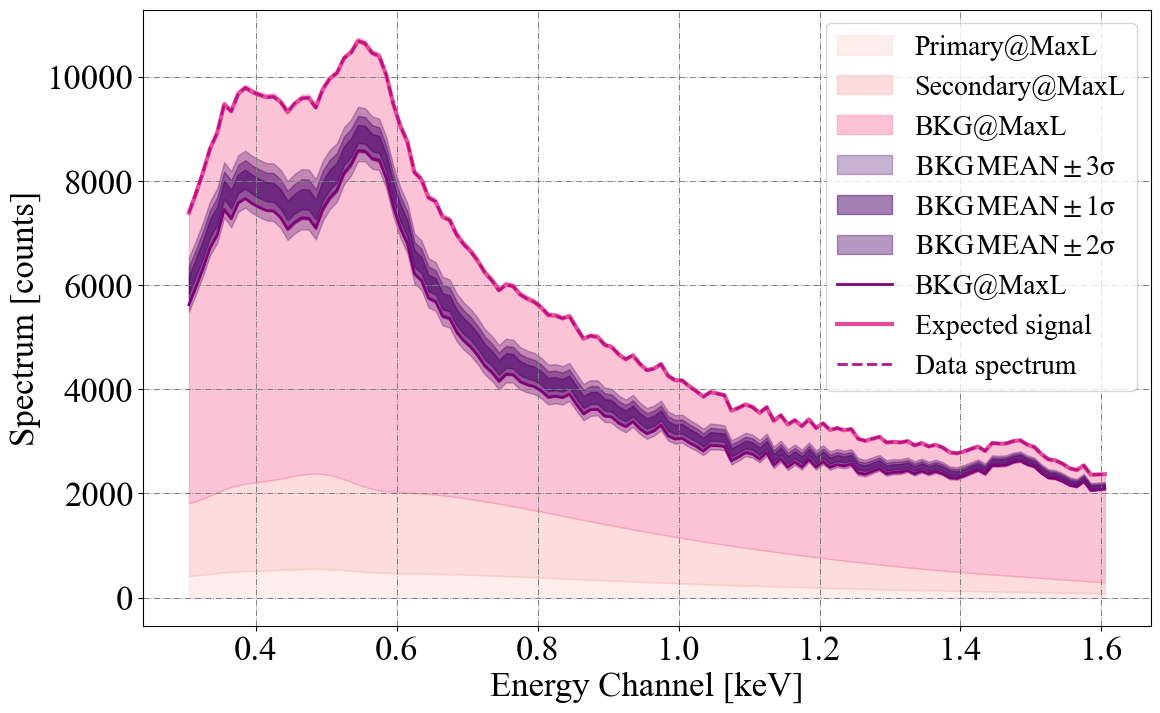}
    \caption{Spectrum for the NICER data with the inferred \texttt{ST+CDT} model components for the case of hydrogen (left panel) and helium (right panel) atmospheric compositions, as well as the associated background. The dashed magenta line corresponds to the NICER phase-averaged spectrum (Data spectrum in the legend). The solid pink line corresponds to the total expected signal. 
    The solid purple line shows the maximum-likelihood (MaxL) background (BKG; background that maximizes the likelihood for the maximum likelihood sample), and the purple shaded regions (from darker to lighter) show the $\pm1$, 2, and 3 standard deviations (calculated from 200 randomly selected posterior samples). 
    The pink shaded regions, from lighter to darker, show the contributions associated to the maximum-likelihood model for the primary (\texttt{ST}) and  secondary hot spot (\texttt{CDT}), as well as the background.}
    \label{fig:background}
\end{figure*}

\subsection{Hot spot geometry}

\label{sec:geometries}
Figure~\ref{fig:geometries} shows the geometric configurations for the surface patterns discussed in the previous sections. Note that these configurations correspond to the maximum likelihood parameters, and may not be representative of the full posteriors (available in Zenodo; see also Figure~\ref{fig:full_posteriors_hydrogen} and Figure~\ref{fig:full_posteriors_helium} for our preferred models), particularly when they exhibit bimodalities. 
The inferred configurations reveal some differences between the hydrogen and helium atmospheric compositions. 

On the one hand, for the hydrogen composition (panels A, B, and C), the location of the hot spots shows  
some variability depending on the complexity of the model. In particular, in panel (B), the preferred \texttt{ST+CDT} hydrogen model (although with spots located in different hemispheres) shows a strong deviation from antipodality, potentially indicating an off-centered dipolar magnetic configuration. 
The \texttt{CDT} spot is located in the northern hemisphere and is characterized by a hot ($\log_{10}(T_s/{\rm K})=6.18$) and small superseding ($\zeta_s=0.02$~rad) region surrounded by a cold ($\log_{10}(T_{c,s}/{\rm K})=5.38$) and large ($\zeta_{c,s}=1.55$~rad) ceding region\footnote{For comparison, the temperature of the ceding region is similar to that derived for the \texttt{elsewhere} component of the \texttt{ST-U+EL} hydrogen model: $\log_{10}(T_{\rm{else}}/\rm{K}) = 5.41\pm0.04$.}, covering approximately half of the surface. The \texttt{ST} spot is located in the southern hemisphere and is slightly larger ($\zeta_p=0.04$~rad) and  hotter ($\log_{10}(T_p/{\rm K})=6.22$) than the superseding region of the \texttt{CDT} spot. However, as previously mentioned, the values reported above and plotted in Figure~\ref{fig:geometries} correspond to the maximum likelihood parameters. In fact, by looking at the full posterior distributions (Figure~\ref{fig:full_posteriors_hydrogen} in Appendix B), they reveal a bimodality for $\zeta_{c,s}$, with a secondary solution around zero, which results in a poorly constrained parameter $\zeta_{c,s}= 0.8^{+0.6}_{-0.7}$~rad.

On the other hand, for the helium composition (panels D, E, and F), the hot spots show fairly similar locations for the different model complexities. In particular, in panel (E), the preferred \texttt{ST+CDT} helium model shows only a slight deviation from antipodality, suggesting a magnetic field configuration closer to a centered dipole. Again, the \texttt{CDT} spot is located in the northern hemisphere and the \texttt{ST} spot in the southern hemisphere. However,  the ceding region of the \texttt{CDT} spot is substantially smaller ($\zeta_{c,s}=0.11$~rad) than in the hydrogen case, forming only a thin ring surrounding the superseding spot. The temperature of the ceding region is colder\footnote{The temperature of the ceding region is slightly higher than  that derived for the \texttt{elsewhere} component of the \texttt{ST-U+EL} helium model: $\log_{10}(T_{\rm{else}}/\rm{K}) = 5.36^{+0.06}_{-0.11}$.} ($\log_{10}(T_{c,s}/{\rm K})=5.61$) than that of the superseding region ($\log_{10}(T_{s}/{\rm K})=6.18$). As in the hydrogen case, the \texttt{ST} spot is smaller ($\zeta_p=0.04$~rad)  and  slightly hotter ($\log_{10}(T_p/{\rm K})=6.28$)  than the superseding component of the \texttt{CDT} spot. By checking the full posterior distributions (Figure~\ref{fig:full_posteriors_helium} in Appendix B), we notice that the maximum likelihood geometries reported for the case of helium in Figure~\ref{fig:geometries} should be taken with care, as clear bimodalities are present in the $\phi_p$, $\phi_s$ and $\Theta_s$ parameters. In any case, given the stability on the geometries shown in panels D, E, and F (i.e., nearly independent of the surface pattern complexity), our results appear to be robust for the helium atmospheric composition. 

\subsection{NICER background}
Figure~\ref{fig:background} shows the contribution to the NICER spectrum from the background as well as the hot spots, when considering \texttt{ST+CDT} model for hydrogen and helium atmospheric compositions. A large background contribution is inferred to be present in the data (for comparison with the bright pulsar PSR~J0030$+$0451, see, e.g., Figure~5 in \citealt{Vinciguerra2024}), representing approximately  $82\%$ and $77\%$ of the NICER counts for the case of models with hydrogen and helium atmospheric composition, respectively (note that the inferred distribution medians for the background are fairly similar between hydrogen and helium models, but the maximum likelihood background used to compute these percentages show some deviations between the two cases). This is consistent with the $\sim80\%$ background reported  by \cite{Bogdanov2019a} in a preliminary analysis of PSR~J2124$-$3358.  In addition, relatively large uncertainties in the background contribution  are present, likely due to the fact that the NICER background constraints are derived from Chandra archival data with a relatively low count statistic for the source. 

Interestingly, Figure~\ref{fig:background} (left panel) also illustrates the spectral effect of the ceding component of the \texttt{CDT} model in the hydrogen case. For the maximum likelihood sample, this component covers nearly half of the stellar surface with a temperature of $\sim 10^{5.4}$~K (see again Figure~\ref{fig:geometries}, panel B, and discussion in  Section~\ref{sec:geometries}).  
The peak of the spectrum associated with this cold component lies below the NICER energy band, and therefore only the Wien tail of the distribution is visible. This manifests as a small spectral contribution in the $\sim 0.3-0.4$~keV range compared to the helium-atmosphere case (Figure~\ref{fig:background}, right panel), for which we infer a comparable temperature but over a substantially smaller ceding area.

\section{Discussion}
\label{sec:discussion}

Our PPM favors of a helium atmospheric composition for the surface of PSR J2124$-$3358. Although this source is an isolated MSP, the only known evolutionary channel for the formation of MSPs and acceleration to millisecond rotational periods involves accretion from a binary companion \citep{Alpar1982,Radhakrishnan1982}. This implies that PSR~J2124$-$3358 was formerly part of a binary system and that its companion was subsequently evaporated, likely  due to the energetic pulsar wind \citep{vandenHeuvel1988,Ruderman1989}. Given that a substantial fraction of MSPs are indeed found in binary systems with  white dwarf companions \citep[e.g.,][]{Manchester2017}, a plausible scenario is that the former companion of  PSR J2124$-$3358 was a helium white dwarf or hydrogen-depleted star, whose accretion and subsequent  evaporation may have led to the deposition of helium onto the NS surface. Although not decisively, this result represents substantial evidence for an MSP with a helium atmospheric composition. 

We note, however, that PSR~J2124$-$3358 is not the first MSP for which a helium atmospheric composition has been suggested. In fact,  analyses of NICER data of PSR~J1231$-$1411 have also shown that its X-ray emission is better explained by a helium atmosphere \citep{Salmi2024b}. Interestingly, PSR~J1231$-$1411 is part of a binary system with a white dwarf companion possessing a hydrogen envelope \citep{Bassa2016}. Since the atmospheric composition of PSR~J1231$-$1411 is therefore unlikely to be explained by past accretion from the companion, \cite{Salmi2024b} considered diffuse nuclear burning \citep[e.g.,][]{Wijngaarden2019} as a possible explanation for the inferred atmospheric composition of that source. This interpretation is particularly interesting because it raises broader questions regarding the atmospheric compositions of other MSPs observed by NICER. Given that the timescale for diffuse nuclear burning is relatively short \citep[$\sim 10^3$~yr,][]{Chang2003,Chang2004} compared to the typical ages of MSPs ($\sim10^9$~yr), and assuming no additional mechanisms are at work, it remains unclear how MSPs such as PSR~J0437$-$4715 can still preferentially exhibit hydrogen atmospheric compositions \citep[see e.g.,][]{GonzalezCaniulef2019,Miller2026}. Future PPM analyses of MSPs observed by NICER will help clarify whether helium atmospheres are uncommon exceptions or a more widespread feature among these sources.

\begin{figure}[t]
    \centering
    \includegraphics[width=\linewidth]{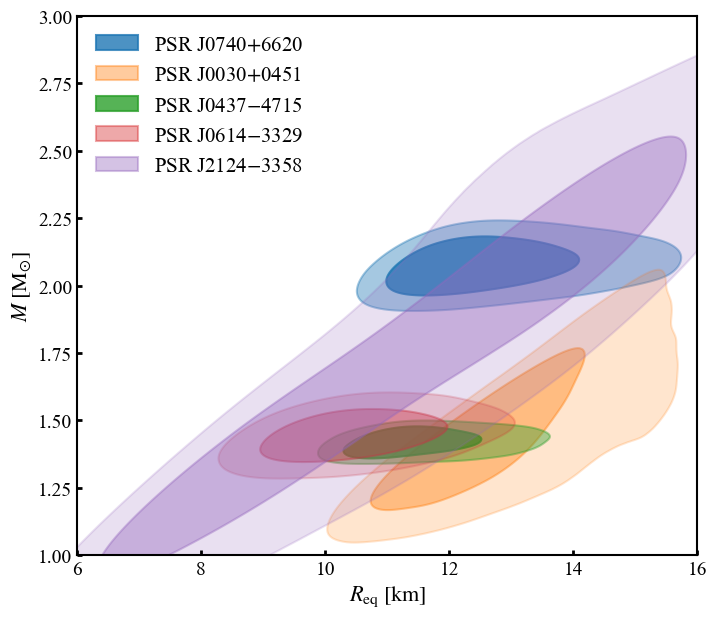}
    \caption{Mass-radius constraints from NICER observations of MSPs. The filled contours show the $68\%$ and $95\%$ credible regions. For PSR~J2124$-$3358, we present the results inferred with the preferred \texttt{ST+CDT} model considering helium atmospheric composition.  We also show the latest published constraints derived with X-PSI for PSR~J0740$+$6620 \citep{Salmi2024a}, PSR~J0030$+$0451 \citep{Kini2026}, PSR~J0437$-$4715 \citep{Choudhury2024}, and PSR~J0614$-$3329 \citep{Mauviard2025}. The tighter mass-radius constraints for PSR~J0740$+$6620, PSR~J0437$-$4715, and PSR~J0614$-$3329 are due in part to the fact that they belong to binary systems, which provide informative priors on their mass and inclination.}
    \label{fig:MR_constraints}
\end{figure}

In addition, our PPM returned stable mass and radius measurements for 
the helium composition (fairly consistent results are obtained for different surface pattern models, see posterior distributions of $M$, $R_{\rm{eq}}$, and compactness in Figure~\ref{fig:combined_posteriors}, right panel).
In fact, for the preferred \texttt{ST+CDT} helium model,  the median value of the radius $R_\mathrm{eq}=11.7$\,km is in good agreement with previous NICER analyses of other MSPs, and the median value of the mass  $M=1.8\,M_\odot$  suggests a relatively massive MSP.  However, since the source is isolated (no independent mass measurement is available), faint, and background-dominated in the NICER observations, we obtained broad posterior distributions, which result in relatively large
credible intervals, for example, as large as $\approx \pm 3$\,km for the radius and  $\approx \pm 0.5\,M_\odot$ for the mass (see Figure~\ref{fig:MR_constraints} for a comparison with mass-radius constraints derived for other MSPs observed by NICER).  If, instead, the actual atmospheric composition were hydrogen (again for \texttt{ST+CDT} model), then the radius and mass medians would be reduced by $\sim 1$\,km and $\sim 0.5\,M_\odot$, respectively. Given the uncertainties, we leave detailed EOS investigations for future studies of this source.

Furthermore, for the hot spots with helium composition, we obtained only slight deviations from antipodality. This suggests that the magnetic field configuration is relatively close to a centered dipole. In contrast, in the case of hydrogen, a substantial off-centered dipole configuration is inferred from our PPM. Future multiwavelength analyses incorporating radio and gamma ray emission, which can also help to constrain the geometry of MSPs \citep[e.g.,][]{Bilous2019,Chen2020,Petri2023,Petri2025,Petri2026}, will provide further insight into the magnetic topology of this source and consequently support for the  helium or hydrogen atmospheric composition.

Motivated by the potential presence of a cold thermal component \citep[like that in PSR~J0437$-$4715,][]{GonzalezCaniulef2019}, we also studied a configuration consisting of a \texttt{ST-U} plus an \texttt{elsewhere} component, that is, including emission from the rest of the NS surface\footnote{Indeed, although MSPs are old objects, thermal emission from the bulk of the MSP surface is theoretically expected \citep[see e.g.,][]{Rodriguez2026}, as they may still be heated by internal mechanisms driven by spin-down, such as vortex creep \citep{Alpar1984} and chemical imbalances \citep[also known as ``rotochemical heating'',][]{Reisenegger1995}.}. We found that this \texttt{elsewhere} component is not needed or contributes only minimally for either a hydrogen or helium composition.  However, for the \texttt{ST+CDT} configuration with a hydrogen composition, the ceding part of the secondary spot covers approximately half of the surface. In fact, this ceding component is relatively cold ($\log_{10} (T_{c,s}/\rm K) \approx 5.4$), and the Wien tail of the spectrum may contribute to the X-ray range. Therefore, observational studies broadening to lower energies, to better characterize the relatively cold surface emission, or with future missions like the enhanced X-ray Timing and Polarimetry mission \citep[eXTP,][]{Zhang2025} will be crucial for better constraining the properties of PSR~J2124$-$3358.

Finally, there are some caveats associated with the analysis presented in this work. 
On the one hand, for the surface patterns assuming a hydrogen atmospheric composition, the peak of the mass posterior distributions reaches the $1~M_\odot$ lower boundary of the prior \citep[limit of validity for the oblateness approximation used in X-PSI,][]{AlGendy2014}, suggesting that the reported masses are likely affected by the truncation of the posterior. Similarly, the inferred radii may also be affected, given their correlation with the mass.
This problem is not present in the case of a helium atmospheric composition, where the mass and radius posteriors are well centered within the prior bounds. However, the helium case does exhibit mild bimodalities in the mass and radius posterior distributions. A more thorough exploration should therefore consider \textsc{MULTINEST} sampling with multimodal exploration turned on, which would allow a more complete characterization of the origin of these bimodalities. In addition, we explored different surface patterns assuming fully ionized atmospheres. This assumption may require relaxation, since the cold components of the \texttt{ST+CDT} and \texttt{ST-U+EL} models produce temperatures of $\sim 10^{5.5}$~K, for which partially ionized atmosphere models should be considered. In any case, for the preferred \texttt{ST+CDT} model with helium composition, the cold ceding region of the \texttt{CDT} spot is relatively small, and therefore its contribution within the NICER energy band is likely small.

All in all, we have presented a significant advance in the study of PSR~J2124$-$3358, but further investigations are still required to derive meaningful and robust constraints on the EOS of superdense matter.

\section{Conclusions}
\label{sec:conclusions}

We performed PPM of PSR~J2124$-$3358 using NICER and Chandra observations, exploring different hot spot complexities and atmospheric compositions. Our analysis provides a substantial evidence in favor of a helium atmospheric composition in an MSP, which yields mass and radius measurements consistent with previous NICER results for other MSPs: $M= 1.8\pm0.5\,M_\odot$  and $R_\mathrm{eq} =11.7^{+2.6}_{-3.0}$\,km (the broad uncertainties are primarily due to the faint nature of the source, the background-dominated observations by NICER, and the lack of an independent mass measurement). The inferred hot-spot geometries suggest a magnetic field configuration close to a centered dipole in the helium case, while a more significantly off-centered dipole is preferred under the hydrogen assumption. 

Although current measurements are not yet sufficiently tight to place strong constraints on the EOS, this study represents a significant step forward in understanding the properties of PSR~J2124$-$3358. Future multiwavelength analysis, incorporating independent constraints for the  magnetic field configuration from radio and gamma-ray analyses, will be essential to further confirm the atmospheric composition, determine the thermal distribution, and emission geometry of this system, ultimately enabling more stringent constraints on the EOS of superdense matter.

\begin{acknowledgments}
D.G.C., S.G., P.S., L.M., and C.K. are supported by the CNES, and by the ANR (Agence Nationale de la Recherche) through the grants ANR-20-CE31-0010 (ANR MORPHER) and ANR-25-CE31-7901-01 (ANR DENSER). A.L.W., M.H. and Y.K. acknowledge support from NWO grant ENW-XL OCENW.XL21.XL21.038 \textit{Probing the phase diagram of Quantum Chromodynamics} (PI: Watts). B.D., D.C. and A.L.W. acknowledge support from European Research Council (ERC) Consolidator Grant (CoG) No. 865768 AEONS (PI: Watts). T.S. acknowledges support by the Research Council of Finland grant No. 368807 and the Centre of Excellence in Neutron-Star Physics (project 374063). This work was performed using HPC resources from CALMIP (project p19056).  We acknowledge NWO for providing access to Snellius, hosted by SURF through the Computing Time on National Computer Facilities call for proposals. The Nan{\c c}ay radio Observatory is operated by the Paris Observatory, associated with the French Centre National de la Recherche Scientifique (CNRS), and partially supported by the Region Centre in France. We acknowledge financial support from the ``Action Th\'ematique de Cosmologie et Galaxies'' (ATCG), ``Action Th\'ematique Gravitation R\'ef\'erences Astronomie M\'etrologie'' (ATGRAM) and ``Action Th\'ematique Ph\'enomènes Extr\^emes et Multi-messagers'' (ATPEM) of CNRS/INSU, France.

\end{acknowledgments}





%
\facilities{NICER, Chandra, NRT}

\software{astropy \citep{astropy2013,astropy2018,astropy2022}, CIAO \citep{Fruscione2006,Fruscione2026}, 
Cython \citep{Behnel2011}, fgivenx \citep{Handley2018}, GetDist \citep{Lewis2025}, GNU Scientific Library \citep{Galassi2009}, HEASoft \citep{heasoft2014}, IPython \citep{Perez2007}
Matplotlib \citep{Hunter2007}, Jupyter \citep{Kluyver2016},
MPI for Python \citep{Dalcin2008}, MultiNest \citep{Feroz2009}, nestcheck \citep{Higson2018}, NumPy \citep{vanderWalt2011}, 
PyMultiNest \citep{Buchner2014}, Python/C language \citep{Oliphant2007}, SciPy \citep{Jones2001}, X-PSI \citep{xpsi2023}.}


\appendix

\section{Residuals}

\begin{figure*}
    \centering
    \includegraphics[width=0.49\linewidth]{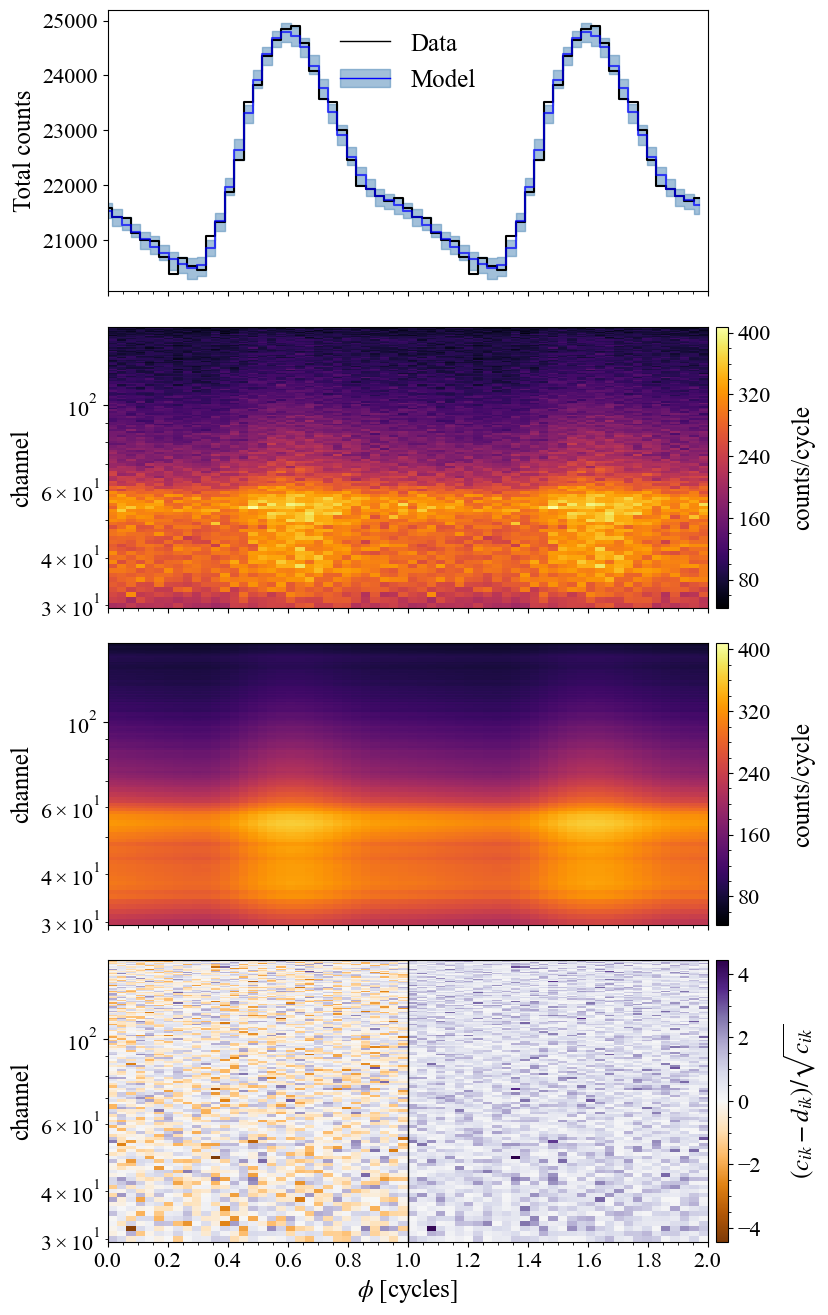}
    \includegraphics[width=0.49\linewidth]{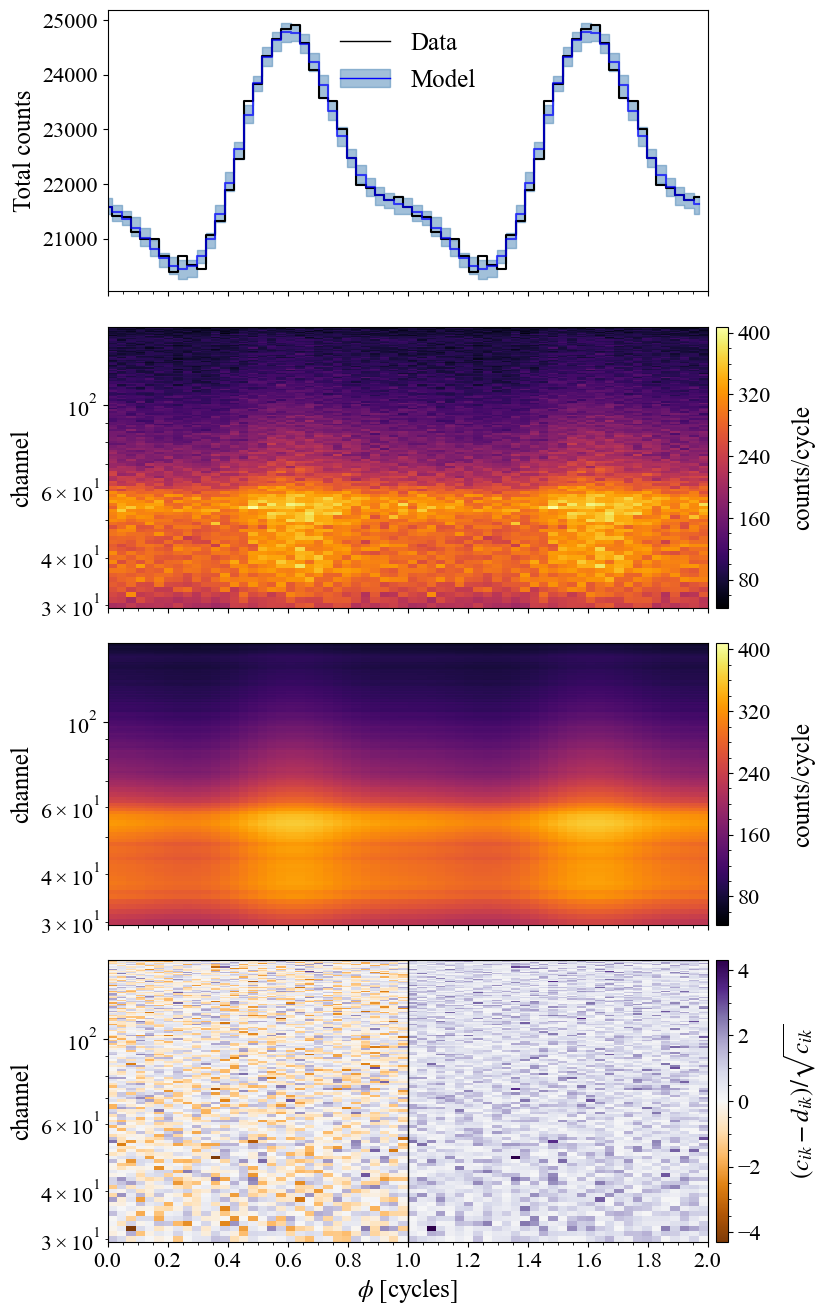}
    \caption{NICER data versus model for the \texttt{ST+CDT} surface pattern considering hydrogen (left column) and helium (right column) atmospheric composition.  For each column, the top panel shows the observed, bolometric pulse profile (integrated over energy channels; black line) together with the corresponding model (blue line), with the shaded region indicating the $68\%$ credible interval. The second panel shows the observed phase- and channel-dependent pulse profile (as shown in Figure~\ref{fig:nicer_data}). The  third panel  shows the phase- and channel-dependent pulse profile model. The bottom, left (right) sub-panel shows the residuals, $(c_{ik}-d_{ik})/\sqrt{c_{ik}}$ (and their absolute values), where $d_{ik}$ and $c_{ik}$ correspond to the data and model counts, respectively, in the $i$-th rotational phase bin and $k$-th channel.}
    \label{fig:residuals}
\end{figure*}

Figure~\ref{fig:residuals} shows the NICER data versus model, as well as the corresponding residuals, for hydrogen and helium atmospheric compositions in the case of the preferred \texttt{ST+CDT} hot spots. The residuals are similar for both atmospheric compositions and show no significant phase- or energy-dependent structures, suggesting that both models provide a good description of the data.

\section{Posteriors and summary tables}

\begin{figure*}
    \centering
    \includegraphics[width=\linewidth]{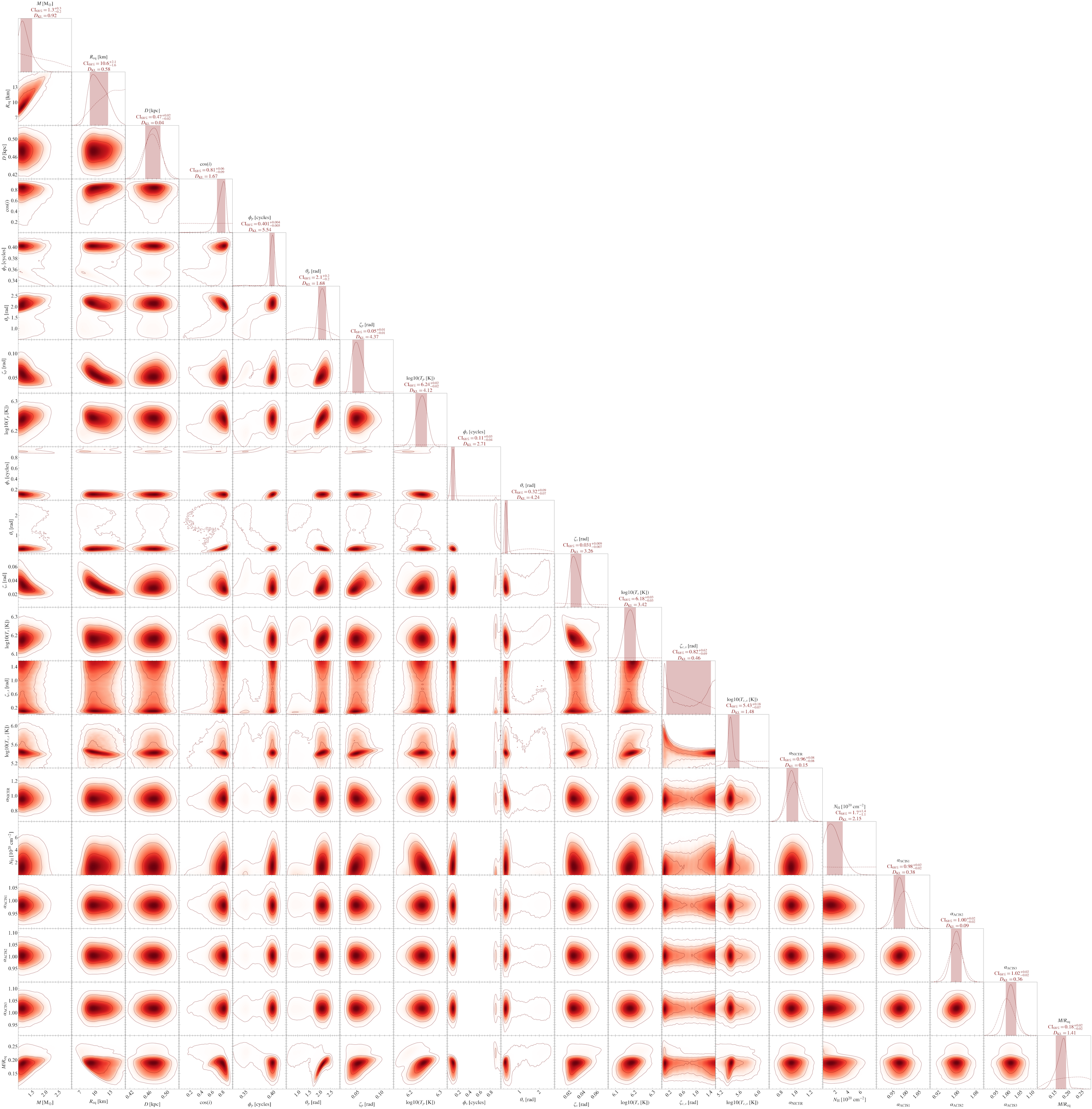}
    \caption{Posterior distributions for the \texttt{ST+CDT} model parameters considering a hydrogen atmospheric composition. In the diagonal panels, the dashed lines show the prior distributions, the solid lines show the posterior distributions,  and the shaded vertical bands indicate the $68.3\%$ credible interval. The resulting model parameters are reported as the median values with corresponding $1\sigma$ uncertainties (the Kullback–Leibler divergence, $D_{\mathrm{KL}}$, is also provided). The contours in the 2D posterior distributions show the $68.3\%$, $95.4\%$, and $99.7\%$ credible regions.}
    \label{fig:full_posteriors_hydrogen}
\end{figure*}

\begin{figure*}
    \centering
    \includegraphics[width=\linewidth]{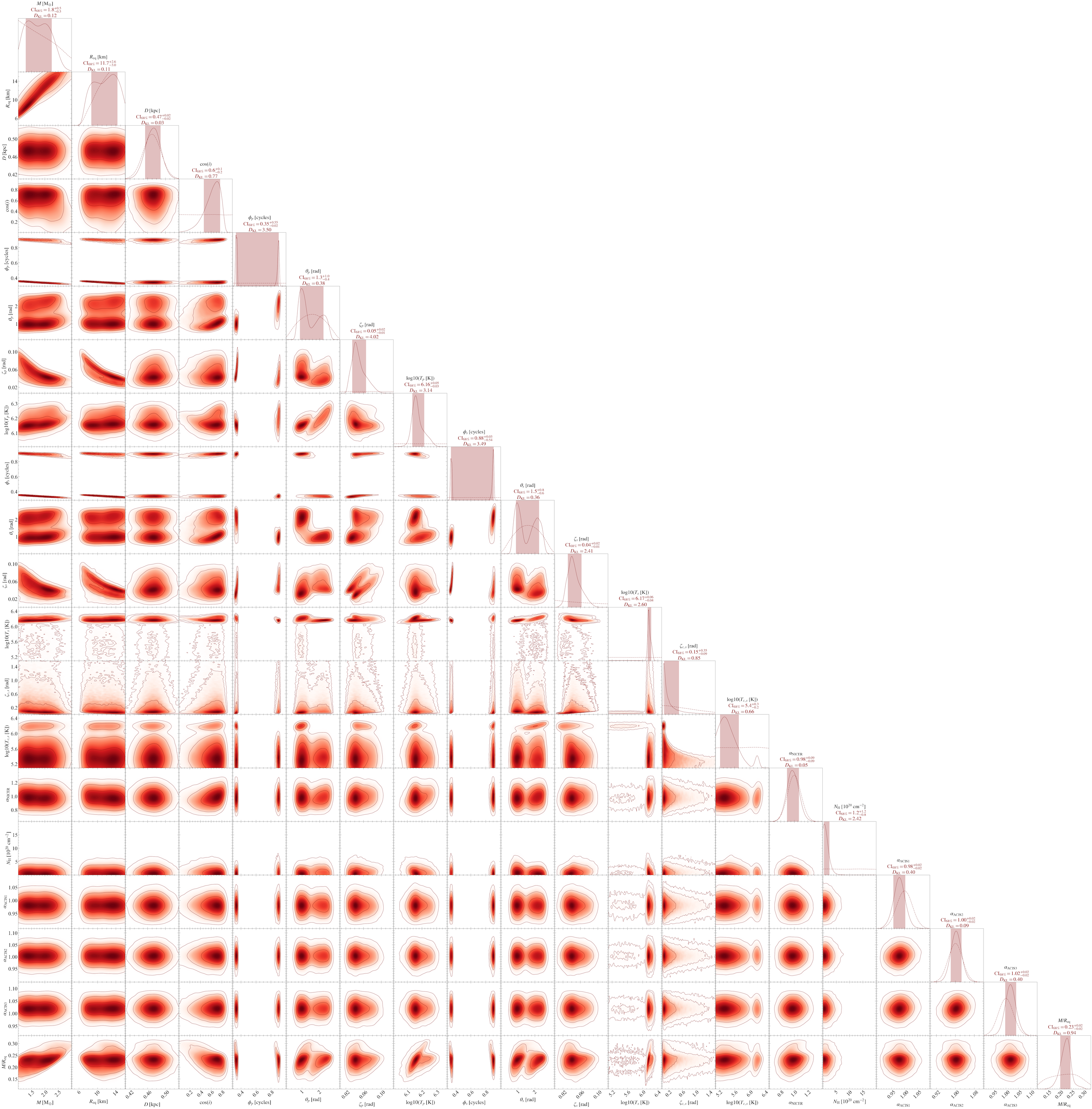}
    \caption{Same as Figure~\ref{fig:full_posteriors_hydrogen}, but for helium atmospheric composition.}
    \label{fig:full_posteriors_helium}
\end{figure*}

Here, we present the full posterior distributions of the preferred \texttt{ST+CDT} model obtained for hydrogen (Figure~\ref{fig:full_posteriors_hydrogen}) and helium atmospheric compositions (Figure~\ref{fig:full_posteriors_helium}), as well as a summary in Table~\ref{tab:full_parameters}. Higher-resolution figures of the full corner plots can be found in the Zenodo repository.

\begin{table*}[!ht]
    \centering
    \caption{Summary for the \texttt{ST+CDT} models considering hydrogen and the preferred helium atmospheric compositions.}
    \begin{tabular}{llrrrrrr}
    \hline \hline 
         &  & \multicolumn{3}{c}{Hydrogen} &\multicolumn{3}{c}{Helium}\\
        \cmidrule(lr){3-5} \cmidrule(lr){6-8}
         Parameter &  Prior PDF & $\widehat{\rm CI}_{68\%}$ & $\widehat{\rm ML}$ & $\widehat{D}_{\rm KL}$  & $\widehat{\rm CI}_{68\%}$ & $\widehat{\rm ML}$ & $\widehat{D}_{\rm KL}$ \\
        \hline
        $P$ [ms] &  4.9 (fixed) & ... & ... & ... & ... & ... & ... \\
        \hline
        $M$ [$M_\odot$] &  $\mathcal{U}(1,3)$ & $1.3^{+0.3}_{-0.2}$ &  1.073 & 0.92 & $1.8\pm0.5$ &  2.489 & 0.12 \\
        $R_{\rm eq}$ [km] &  $\mathcal{U}(3r_g(1), 16)$ & $10.6^{+2.1}_{-1.6}$ & 12.31 & 0.58   & $11.7^{+2.6}_{-3.0}$  & 15.14 & 0.11\\        
        $D$ [kpc] &  $\mathcal{N}(0.470,0.02^2)$ & $0.47\pm0.02$ & 0.472 & 0.04 & $0.47\pm0.02$ & 0.482  &0.03\\
        $\cos(i)$&  $\mathcal{U}(0,1)$ & $0.81^{+0.06}_{-0.09}$ & 0.862 & 1.67   & $0.65^{+0.12}_{-0.18}$ & 0.699 & 0.77\\  
        $\phi_p$ [cycles]  &  $\mathcal{U}(0,1)$ & $0.401^{+0.004}_{-0.005}$ &  0.399 & 5.54   & $0.35^{+0.55}_{-0.02}$ & 0.898  & 3.50\\
        $\theta_p$ [rad]  &  $\cos(\theta_p) \sim \mathcal{U}(-1,1)$ & $2.13\pm0.18$ & 1.803 & 1.68  & $1.27^{+0.95}_{-0.39}$ &  2.739 & 0.38\\
        $\zeta_p$ [rad]  &  $\mathcal{U}(0,\pi/2 )$ & $0.05\pm0.01$ & 0.040 & 4.37  & $0.05^{+0.02}_{-0.01}$ & 0.043 & 4.02\\
        $\log_{10}(T_p \text{[K]})$ &  $\mathcal{U}(5.1,6.5)$ & $6.24\pm0.02$ & 6.221 & 4.12  & $6.16^{+0.05}_{-0.02}$ & 6.282 & 3.14\\
        $\phi_s$ [cycles]  &  $\mathcal{U}(0,1)$ & $0.11^{+0.03}_{-0.04}$ & 0.107 & 2.71  &  $0.88^{+0.03}_{-0.54}$ & 0.324 &3.49\\
        $\theta_s$ [rad]  &  $\cos(\theta_s) \sim \mathcal{U}(-1,1)$ & $0.32^{+0.09}_{-0.07}$ & 0.415 & 4.24  & $1.47^{+0.78}_{-0.59}$  & 1.071 & 0.36\\
        $\zeta_s$ [rad]  &  $\mathcal{U}(0,\pi/2 )$ & $0.031^{+0.009}_{-0.007}$ & 0.022 & 3.26  & $0.04^{+0.02}_{-0.01}$  & 0.035 & 2.41\\
        $\log_{10}(T_s \text{[K]})$ &  $\mathcal{U}(5.1,6.5)$ & $6.18\pm0.03$ &  6.179 & 3.42  & $6.17^{+0.06}_{-0.04}$ &  6.185 & 2.60\\
        $\zeta_{c,s}$ [rad]  &  $\mathcal{U}(0,\pi/2)$ & $0.82^{+0.62}_{-0.69}$ & 1.555 & 0.46  & $0.15^{+0.35}_{-0.09}$ & 0.110 & 0.85 \\
        $\log_{10}(T_{c,s} \text{[K]})$ &  $\mathcal{U}(5.1,6.5)$ & $5.43^{+0.18}_{-0.07}$ & 5.384 & 1.48  & $5.41^{+0.30}_{-0.20}$ & 5.609 & 0.66\\
        $N_{\rm H} [10^{20} \text{cm}^{-2}]$ &  $\mathcal{U}(0.001,20)$ & $1.7^{+1.4}_{-1.1}$  & 0.324 & 2.15  &  $1.19^{+1.25}_{-0.81}$& 0.347 &2.42\\
        $\alpha_{\rm{NICER}} $ &  $\mathcal{N}(1.0,0.10^2)$ & $0.96\pm0.08$ & 0.858 & 0.15  & $0.98\pm0.09$ &  1.148 & 0.05\\
        $\alpha_{\rm{ACIS1}} $ &  $\mathcal{N}(1.0,0.03^2)$ & $0.98\pm0.02$ & 0.962 & 0.38  & $0.98\pm0.02$ &  0.922 & 0.40\\
        $\alpha_{\rm{ACIS2}} $ &  $\mathcal{N}(1.0,0.03^2)$ & $1.00\pm0.02$  & 1.039 & 0.09 & $1.00\pm0.02$ &  0.990 & 0.09\\
        $\alpha_{\rm{ACIS3}} $ &  $\mathcal{N}(1.0,0.03^2)$ & $1.02\pm0.02$ & 1.057 & 0.36 & $1.02\pm0.02$ &  1.029 & 0.40\\
        \hline \hline
        & Sampling information  & \multicolumn{3}{c}{Hydrogen} & \multicolumn{3}{c}{Helium}\\
        \hline
        & Number of free parameters:  & \multicolumn{3}{c}{19} & \multicolumn{3}{c}{19}\\
        & Number of live points: & \multicolumn{3}{c}{$10^4$} & \multicolumn{3}{c}{$10^4$}\\
        & Sampling efficiency: & \multicolumn{3}{c}{0.1} & \multicolumn{3}{c}{0.1}\\
        & Evidence tolerance: & \multicolumn{3}{c}{0.1} & \multicolumn{3}{c}{0.1}\\
        & Likelihood evaluations: & \multicolumn{3}{c}{$405,638,985$} & \multicolumn{3}{c}{$1,784,868,990$}\\
        & Estimated evidence ($\ln \mathcal{Z}$):& \multicolumn{3}{c}{$-22110.530\pm0.061$} & \multicolumn{3}{c}{$-22108.579\pm0.056$}\\
        & Computation time (CPU hours): & \multicolumn{3}{c}{$30,873$} & \multicolumn{3}{c}{$221,321$}\\
        \hline \hline
    \end{tabular}
    \label{tab:full_parameters}
        \begin{tablenotes}
        \item \textsc{Notes} - The different parameters are explained in Table~\ref{tab:params} (the subscripts $p$ and $s$ denote the parameters associated with the primary and secondary hot regions, respectively). For each of them,  we list the prior PDFs, the median values together with the corresponding $68.3\%$ credible intervals, $\widehat{\rm CI}_{68\%}$, the maximum-likelihood parameter, $\widehat{\rm ML}$, and the Kullback-Leibler divergence, $\widehat{D}_{\rm KL}$, expressed in bits, and considering either hydrogen or helium atmospheric composition. 
        \end{tablenotes}
\end{table*}

\bibliography{bibliography}{}
\bibliographystyle{aasjournalv7}



\end{document}